\documentclass[aps,prd,twocolumn,superscriptaddress,nofootinbib,10pt,preprintnumbers]{revtex4-2}

\pdfoutput=1
\usepackage[utf8]{inputenc}
\usepackage{graphicx,amsmath,amsfonts,amssymb,aas_macros,slashed}
\usepackage{bbold}
\usepackage{datetime}
\usepackage{numprint}
\usepackage{enumitem}
\usepackage{graphicx,color,amsmath}
\usepackage{hyperref}
\usepackage{booktabs}
\usepackage{multirow}
\usepackage{cancel}

\usepackage{rotating}
\allowdisplaybreaks

\newcommand{\keV}{\ensuremath{\mathrm{keV}}}
\newcommand{\MeV}{\ensuremath{\mathrm{MeV}}}
\newcommand{\GeV}{\ensuremath{\mathrm{GeV}}}
\newcommand{\TeV}{\ensuremath{\mathrm{TeV}}}

\renewcommand{\day}{\ensuremath {\mathrm{day}}}

\renewcommand{\arraystretch}{1.5}

\newcommand{\LambdaUV}{\Lambda}

\usepackage{pgfplots}
\usepackage{bm}
\usepackage{mathtools}

\makeatletter
\newcommand{\thickhline}{%
    \noalign {\ifnum 0=`}\fi \hrule height 1pt
    \futurelet \reserved@a \@xhline
}
\newcolumntype{"}{@{\hskip\tabcolsep\vrule width 1pt\hskip\tabcolsep}}
\makeatother

\begin{document}
\preprint{UWThPh 2023-8}
\preprint{IPMU23-0006}

\title{Multipole vector dark matter below the GeV-scale}

\author{Xiaoyong Chu}
\affiliation{Institute of High Energy Physics, Austrian Academy of Sciences, Georg-Coch-Platz 2, 1010 Vienna, Austria}

\author{Junji Hisano}
\affiliation{
Department of Physics, Nagoya University, Furo-cho Chikusa-ku, Nagoya 464-8602 Japan}
\affiliation{Kobayashi-Maskawa Institute for the Origin of Particles and the Universe, Nagoya University,
Furo-cho Chikusa-ku, Nagoya 464-8602 Japan}
\affiliation{Kavli IPMU (WPI), UTIAS, The University of Tokyo, Kashiwa 277-8584, Japan}

\author{Alejandro Ibarra}
\affiliation{Physik-Department, Technische Universit\"at M\"unchen, James-Franck-Straße, 85748 Garching, Germany}

\author{Jui-Lin Kuo}
\affiliation{Department of Physics and Astronomy,
University of California, Irvine, CA 92697-4575, USA}

\author{Josef Pradler}
\affiliation{Institute of High Energy Physics, Austrian Academy of Sciences, Georg-Coch-Platz 2, 1010 Vienna, Austria}
\affiliation{University of Vienna, Faculty of Physics, Boltzmanngasse 5, A-1090 Vienna, Austria}
\affiliation{CERN, Theoretical Physics Department, 1211 Geneva 23, Switzerland}

\begin{abstract}
We consider electrically neutral complex vector particles $V$ below the GeV mass scale that, from a low energy perspective, couple to the photon via higher dimensional form factor interactions. We derive ensuing astrophysical constraints by considering the anomalous energy loss from the Sun, Horizontal Branch, and Red Giant stars as well as from SN1987A that arise from vector pair-production in these environments. Under the assumption that the dark states $V$ constitute dark matter, the bounds are then complemented by direct and indirect detection as well as cosmological limits. The relic density from freeze-out and freeze-in mechanisms is also computed. On the basis of a UV-complete model that realizes the considered effective couplings, we also discuss the naturalness of the constrained parameter space, and provide an analysis of the zero mass limit of~$V$.
\end{abstract}

\maketitle

\section{Introduction}
New spin-1/2 particles are arguably the most entertained option when considering models of Dark Matter (DM), only to be followed by scalar DM candidates. Vector particles $V^\mu$, on the other hand,  are a comparatively less studied possibility for DM. The reasons for it may be circumstantial rather than fundamental. In the Standard Model (SM), matter fields are fermions, and spin-1 vector fields chiefly take the role of mediating their gauge interactions. Although gluons dynamically deliver a significant fraction of mass to baryons, we are better familiar with attributing 86\% of the Universe's present mass density to new fields with their mass as a fundamental parameter in the Lagrangian or generated through symmetry breaking involving spin-1/2 (or spin-0) fields. However, there is {\it a priori} no reason to discard the possibility that DM is a fundamental (weakly coupled) massive vector field.  

Complex vector DM has been studied in a number of works before where its coupling to SM was mediated by the tree-level exchange of heavy fermion or scalar mediators~\cite{Servant:2002aq,Cheng:2002ej,Hubisz:2004ft,Birkedal:2006fz,Hambye:2008bq,Hisano:2010yh, Davoudiasl:2013jma, Gross:2015cwa,Karam:2015jta,Choi:2019zeb,Elahi:2019jeo,Abe:2020mph,Nugaev:2020zcv,Elahi:2020urr,Hambye:2021xvd}. More recently, the possibility was investigated that $V^\mu$, albeit electrically neutral, shares a coupling with the photon through electromagnetic multipole moments~\cite{Hisano:2020qkq, Krnjaic:2022wor}. In~\cite{Hisano:2020qkq}, the direct detection phenomenology of electroweak scale heavy $V^\mu$ was then investigated, complementing earlier works of electroweak scale DM of spin-0 and spin-1/2~\cite{Bagnasco:1993st,Pospelov:2000bq,Sigurdson:2004zp,Masso:2009mu,Barger:2010gv,Banks:2010eh,Ho:2012bg,Schmidt:2012yg,Kopp:2014tsa,Ibarra:2015fqa,Sandick:2016zut,Kavanagh:2018xeh}.
Once the mass of DM drops below the GeV-scale, a multitude of new phenomenological considerations come into play. The electromagnetic moments may be explored at the intensity frontier, through flavor physics and precision tests, as well as  in astrophysics, through the anomalous energy loss they induce in stars~\cite{Chu:2018qrm,Chu:2019rok,Chu:2020ysb,Chang:2019xva,Marocco:2020dqu}. 

It is the purpose of this paper to carry over those considerations and chart out the parameter space for sub-GeV electrically neutral complex vector DM candidates that carry electromagnetic (EM) form factor interactions. In the classification of their transformation property under discrete symmetries and dimensionality, those are the mass-dimension five electric and magnetic dipole interactions and dimension six magnetic and electric quadrupole moments, charge radius interaction as well as toroidal and anapole moments. The vector particles are pair produced from off-shell photons via $\gamma^* \to V^\dag V$. We establish the stringent astrophysical constraints from the anomalous energy loss from the Sun, from Horizontal Branch (HB) and Red Giant Branch (RGB) stars as well as from SN1987A. These constraints probe the existence of such dark states irrespective if they constitute the bulk of DM. Assuming that they are DM, we also study its freeze-out and freeze-in production mechanisms, as well as the direct and indirect detection limits. We outline a UV completion of the electromagnetic effective interactions under investigation, and consider the scaling of production rates in the high-energy limit of~$V^\mu$.

The paper is organized as follows. In Sec.~\ref{sec:emvectors} we introduce the effective Lagrangian and form factors together with the amplitude for $V$ pair production. In Sec.~\ref{sec:stars} we compute the energy loss rates in the considered astrophysical environments and derive ensuing limits. In Sec.~\ref{sec:VDM} we compute the vector relic abundance from either freeze-in or freeze-out and derive various  constraints on the model. In Sec.~\ref{sec:UV} we connect the studied effective interactions to a UV-complete model and discuss various points of importance. Conclusions are presented in Sec.~\ref{sec:conclusions}.

\section{Vectors with EM form factors}
\label{sec:emvectors}

\begin{table}[t]
    \centering
    \begin{tabular}{l|c@{\hskip 7pt}c@{\hskip 7pt}c@{\hskip 7pt}c}
    \toprule
    interaction type    & coupling & $C$ & $P$ & $C\! P$ \\
\midrule
 magn.~dipole & $ \mu_V = \dfrac{e}{2 m_V} (\kappa_\Lambda + \dfrac{m_V^2}{\Lambda^2}\lambda_\Lambda) $  & $+1$  & +1 & +1  \\
  elec.~dipole & $ d_V = \dfrac{e}{2m_V} (\tilde\kappa_\Lambda + \dfrac{m_V^2}{\Lambda^2}\tilde\lambda_\Lambda)$ & $+1$   & $-1$ & $-1$\\
 elec.~quadrupole & $Q_V = -\dfrac{e}{m_V^2} (\kappa_\Lambda - \dfrac{m_V^2}{\Lambda^2}\lambda_\Lambda) $ & +1 & +1 & +1 \\
 magn.~quadrupole & $\tilde Q_V = -\dfrac{e}{m_V^2} (\tilde\kappa_\Lambda - \dfrac{m_V^2}{\Lambda^2}\tilde\lambda_\Lambda)$ & +1 &  $-1$ & $-1$ \\
 charge radius & $g_1^A/m_V^2= g_1^\Lambda/\Lambda^2$ & $+1$  &  $+1$ & $+1$ \\
 toroidal moment & $g_4^A/m_V^2= g_4^\Lambda/\Lambda^2$ & $-1$ &  $+1$ & $-1$ \\
 anapole moment & $g_5^A/m_V^2= g_5^\Lambda/\Lambda^2$ & $-1$ &  $-1$ & +1 \\
         \bottomrule
    \end{tabular}
    \caption{Nomenclature of various  effective interactions considered in this work together with their transformation property under charge conjugation ($C$), parity~($P$) and their combination ($C\! P$) 
    with the respective vector intrinsic parity and charge conjugation assignments chosen as $P V^\mu(t,\vec x) P^{-1} = V_\mu(t,-\vec x)$ and $C V^\mu C^{-1} = -  V^{\dag\mu}$ (and equivalent relations for the photon).}
    \label{tab:cpproperties}
\end{table}

The effective Lagrangian of a complex massive vector field~$V^\mu$ with mass $m_V$, which is electromagnetic neutral but interacts with the SM photon $A^\mu$ via electromagnetic (EM) form factors up to mass-dimension~6, can be expressed as~\cite{Gaemers:1978hg,Hagiwara:1986vm,Gounaris:1996rz,Hisano:2020qkq}
\begin{align}
\label{eq:effectiveL}
\dfrac{\mathcal{L}}{e} &= \dfrac{ig_1^\Lambda}{2\LambdaUV^2} \left[  \left(V^\dagger_{\mu\nu} V^\mu - V^{\dagger \mu} V_{\mu\nu} \right) \partial_\lambda F^{\lambda\nu}  - V^{\dagger \mu} V^\nu \Box F_{\mu\nu} \right] \nonumber \\
&+\dfrac{g_4^\Lambda}{\LambdaUV^2} V^\dagger_\mu V_\nu \left( \partial^\mu \partial_\rho F^{\rho\nu} + \partial^\nu \partial_\rho F^{\rho\mu}\right) \nonumber \\
&+ \dfrac{g_5^\Lambda}{\LambdaUV^2} \epsilon^{\mu\nu\rho\sigma} \left( V_\mu^\dagger \overleftrightarrow{\partial_\rho} V_\nu \right) \partial^\lambda F_{\lambda \sigma} \nonumber \\
&+ i \kappa_\Lambda V_\mu^\dagger V_\nu F^{\mu\nu} + \dfrac{i\lambda_\Lambda}{\LambdaUV^2} V_{\lambda\mu}^\dagger V^\mu_{\,\,\,\,\nu} F^{\nu \lambda} \nonumber \\
&+i \tilde{\kappa}_\Lambda V_\mu^\dagger V_\nu \tilde{F}^{\mu\nu} + \dfrac{i\tilde{\lambda}_\Lambda}{\LambdaUV^2} V_{\lambda\mu}^\dagger V^\mu_{\,\,\,\,\nu} \tilde{F}^{\nu \lambda}\,,
\end{align}
where $\LambdaUV$ characterizes the energy scale below which the effective operator approach is valid, with $\LambdaUV \gg m_V$.\footnote{We impose the electric neutrality of~$V$.
For works on milli-charged vector particles, see, \emph{e.g.}, \cite{Nieves:1996ff, Gabrielli:2015hua}. 
}
where the field strength tensors, their duals and other field second derivatives are defined by,
\begin{align*}
&V_{\mu\nu}=\partial_\mu V_\nu -\partial_\nu V_\mu\,, \quad F_{\mu\nu}=\partial_\mu A_\nu -\partial_\nu A_\mu\,, \\ 
&\tilde{F}_{\mu\nu}= \epsilon_{\mu\nu\rho\sigma}{F}^{\rho\sigma}/2\,,\quad \left( V_\mu^\dagger \overleftrightarrow{\partial_\rho} V_\nu \right) = V^\dagger_\mu (\partial_\rho V_\nu) -(\partial_\rho V_\mu^\dagger) V_\nu\,.
\end{align*}
The total antisymmetric tensor follows the convention that $\epsilon_{0123} = - \epsilon^{0123} = 1$. 
Following the convention adopted in \cite{Hagiwara:1986vm}, we define
\begin{align}
\label{eq:definitionDipoleQuad}
 &\mu_V = \dfrac{e}{2 {m_V}} \kappa_\Lambda + \dfrac{e m_V }{2\Lambda^2}  \lambda_\Lambda\,,\quad Q_V = -\dfrac{e}{{m^2_V}} \kappa_\Lambda + \dfrac{e}{\Lambda^2} \lambda_\Lambda\,, \nonumber \\
 &d_V = \dfrac{e}{2{m_V}} \tilde{\kappa}_\Lambda+ \dfrac{e m_V}{2\Lambda^2} \tilde{\lambda}_\Lambda \,,\quad \tilde{Q}_V = -\dfrac{e}{{m^2_V}}  \tilde{\kappa}_\Lambda + \dfrac{e}{\Lambda^2}  \tilde{\lambda}_\Lambda\,,\nonumber \\
  &g_{1,4,5}^A = \dfrac{m_V^2}{\Lambda^2}   g_{1,4,5}^\Lambda  \,,  
\end{align}
corresponding to magnetic dipole, electric quadrupole, electric dipole and magnetic quadrupole in the first two lines, respectively. 
According to the transformation under discrete Lorentz symmetries, we can see $g_1^A$, $g_4^A$, $g_5^A$ as charge radius, toroidal moment and anapole moment, respectively. 
Note that here $g_1^A$, $g_4^A$ and $g_5^A$ are dimensionless coupling constants; $e$ is the electric charge. As we shall see below, the requirement for its validity is that the typical energy scale of the process $\sqrt{s}\ll v_D$, where $v_D$ is the symmetry breaking scale in the UV description that generates the vector mass.  Naive Dimensional Analysis (NDA)~\cite{Manohar:1983md} suggests that the dimensionless constants, $g_{1,4,5}^\Lambda$, $\kappa_\Lambda$, $\tilde{\kappa}_\Lambda$, $\lambda_\Lambda$, and $\tilde{\lambda}_\Lambda$ can be of order $g_D^2/(4\pi)^2$ with $g_D$ being a UV coupling constant of~$V$; see Sec.~\ref{sec:UV} below for a UV example where some of the couplings are of that order while others are further suppressed.

The Lagrangian~\eqref{eq:effectiveL} induces a $A$-$V$-$V$ interaction. Introducing the momentum assignment $A^\mu (k)  \rightarrow V^\alpha (q) + V^{\dagger\beta}(q')$, with $k$  incoming and $q, q'$  outgoing four-vectors, the interactions in~\eqref{eq:effectiveL} assemble themselves in the vertex factor,
\begin{align}
\label{eq:VertexFactor}
&i\Gamma^{\mu\alpha\beta}(k,p) = -\dfrac{ie g_1^A}{2 m_V^2} k^2  p^\mu g^{\alpha\beta}  \nonumber 
\\
&-\dfrac{e g_4^A}{m_V^2} k^2 (k^\alpha g^{\mu\beta} + k^\beta g^{\mu\alpha}) -\dfrac{e g_5^A}{m_V^2} k^2 \epsilon^{\mu\alpha\beta\rho} p_\rho \nonumber \\
&-2i m_V \mu_V \left[k^\alpha g^{\mu\beta} - k^\beta g^{\mu\alpha} +\dfrac{1}{4m_V^2} \left( k^2 g^{\alpha \beta} p^\mu - 2 k^\alpha k^\beta p^\mu \right) \right] \nonumber\\
&-\dfrac{iQ_V}{4} \left( k^2 g^{\alpha \beta} p^\mu - 2 k^\alpha k^\beta p^\mu \right) \nonumber \\
&-\dfrac{id_V}{2m_V} p^\mu \left[ kp\right]^{\alpha\beta} - \dfrac{i\tilde{Q}_V}{4} \left(p^\mu \left[kp \right]^{\alpha\beta} + 4m_V^2 \epsilon^{\mu\alpha\beta\rho}k_\rho \right)
\,,
\end{align}
with $p \equiv q - q'$ and $[kp]^{\alpha\beta} \equiv \epsilon^{\alpha\beta\rho\sigma} k_\rho p_\sigma$. In deriving the vertex factor suitable for Feynman-diagrammatic computation, 
we have imposed Lorentz gauge so that $\partial^\mu A_\mu = 0$ for ${\rm U(1)_{\rm EM}}$ and used $\partial^\mu V_\mu = 0$ for an on-shell massive vector field.

\subsection{Common squared amplitude}

A $V^\dag V$ pair is produced from an off-shell photon of momentum $k$. Therefore, we may find a formulation of the problem that is common to all processes considered in this work,  by dressing this part of the amplitude with the associated  SM-processes that produce $\gamma^*(k)$. The squared amplitude summed over the three polarizations $\lambda$ and $\lambda'$ of the outgoing vectors is hence given by, 
\begin{align}
 \sum\limits_{\lambda,\lambda'} |\mathcal{M}^{\lambda \lambda'}|^2 = D_{\mu\nu} (k) D^*_{\rho \sigma} (k)\mathcal{T}_{\rm SM}^{\mu\rho} \mathcal{T}_{\rm DM}^{\nu\sigma}\,.
\end{align}
Here, $D_{\mu\nu}(k)$ is the photon propagator and $\mathcal{T}_{\rm SM}^{\mu\rho}$ is the SM current giving rise to $\gamma^*(k)$. It is important to note that $D_{\mu\nu}(k) $ receives  finite temperature corrections when stellar production of $V$-pairs is considered (see below).  
The DM squared matrix element reads
\begin{align}
\mathcal{T}_{\rm DM}^{\nu\sigma} &= \Gamma^{\alpha\beta\nu} (\Gamma^{\alpha' \beta' \sigma})^\dagger \sum\limits_\lambda \epsilon_\alpha^\lambda (q) \epsilon^{\lambda *}_{\alpha'} (q) \sum\limits_{\lambda'} \epsilon^{\lambda'}_\beta (q') \epsilon_{\beta'}^{\lambda' *}(q') \nonumber \\
&=  \Gamma^{\alpha\beta\nu} (\Gamma^{\alpha' \beta' \sigma})^\dagger \left(-g_{\alpha \alpha'} + \dfrac{q_\alpha q_{\alpha
'}}{m_V^2}\right)
\left(-g_{\beta \beta'} + \dfrac{q'_\beta q'_{\beta
'}}{m_V^2}\right).
\end{align}
For as long as one is not concerned with the  $V$-differential distributions of energy or angle in the medium or laboratory frame, one may integrate over the phase space $\Phi_2$ of the $V$-pair,  
\begin{align}
\label{eq:fofs}
  I_{\rm DM}^{\nu \sigma} \equiv \int d\Phi_2 \, \mathcal{T}_{\rm DM}^{\nu \sigma} 
 = \dfrac{1}{8\pi}\sqrt{1-\dfrac{4m_V^2}{s}} f(s) \left( -g^{\nu\sigma} + \dfrac{k^\nu k^\sigma}{s} \right).
\end{align}
The entire information of the various form factors is contained in the dimension-2 function $f(s)$,  where $s = k^2$ is the invariant mass of the $V$-pair. The resulting expressions are listed in Tab.~\ref{tab:results}, where we have assumed that operators do not interfere.
The function $f(s)$ feeds into the computed vector production rates below and is hence of central importance. Note that with increasing mass-dimension of the coupling, the power of~$s$ appearing in $f(s)$ increases as well. This provides a UV-biasing of the $V$-production rates. 

\section{Stellar energy loss}
\label{sec:stars}

\begin{table}[t]
\centering
\begin{tabular}{ r|rrr } 
 \toprule
  & $\omega_p$ & $T$ & \text{thermal plasma} \\ 
  \midrule
 Sun's core & 0.3\,{\rm keV} & 1.4\,{\rm keV}  &  \text{classical} \\ 
 HB's core & 2.6\,{\rm keV} & 10.6\,{\rm keV}  & \text{classical}\\ 
  RG's core & 8.6\,{\rm keV} & 8.6 \,{\rm keV} & \text{non-relativistic/degenerate}\\ 
  SN's core & 17.6\,{\rm MeV} & 12.1\,{\rm MeV} & \text{relativistic/degenerate}\\ 
 \bottomrule
\end{tabular}
\caption{Stellar objects considered in this work together with the typical core plasma frequency and photon temperature considered in this work. Here classical regime refers to a non-relativistic ($T \ll m_e$) and non-degenerate ($T \gg \mu_e - m_e$) plasma, where $\mu_e$ and $m_e$ are the electron chemical potential and mass. }
\label{tab:environment}
\end{table}

In this section, we derive constraints on the EM form factors of~$V$ from stellar energy loss. 
We follow~\cite{Chu:2019rok} for details on  stellar environments, $V$-production processes, and $V$-trapping in SN.
A summary of temperature and plasma frequency of each environment can be found in Tab.~\ref{tab:environment}.

\subsection{RG, HB stars and the Sun}

The anomalous energy loss induced by $V$ pair production and subsequent escape can be constrained by  observations of lifetime and relative composition of stars.
For RG stars, we impose that the energy loss rate in the stellar core, $\dot{Q}_{\rm RG}$, should not exceed 
\begin{align}
   \dot{Q}_{\rm RG} < 10 \,{\rm erg/g/s} \times \rho_{\rm RG}\, .
\end{align}
This criterion is obtained by disallowing an increase in  core mass prior to helium ignition by more than $5\%$~\cite{1996slfp.book.....R}.
For the energy density and photon temperature of the core we adopt $\rho_{\rm RG} = 2\times 10^5\ {\rm g/cm^3}$ and $T = 8.6\ \keV$, respectively.
Energy loss carried by $V$ also changes the helium-burning lifetime in HB stars, causing an imbalance of the stellar RG vs.~HB star population in globular clusters. 
A conservative constraint on non-standard energy loss in HB stars reads~\cite{1996slfp.book.....R}
\begin{align}
\int_{\rm core} dV \, \dot{Q}_{\rm HB} < 10\% \times L_{\rm HB}\,,
\end{align}
where we take $L_{\rm HB}  = 20 L_{\odot}$ for a $0.5\,M_{\odot}$ core with $L_{\odot} = 3.83 \times 10^{33} \,{\rm erg/s}$ and $M_{\odot} = 1.99 \times 10^{33} \,{\rm g}$ being the Solar luminosity and Solar mass.
For the Sun, a benchmark criterion can be drawn from total Solar photon luminosity~\cite{Frieman:1987ui,PhysRevD.40.942},
\begin{align}
\label{eq:SunCriterion}
   \int_{\rm Sun} dV\,\dot{Q}_{\odot} < 10\% \times L_\odot\,.
\end{align}
We note that a more stringent criterion is possible, see, e.g.,~\cite{1998SSRv...85..161G,2009ARA&A..47..481A,Redondo:2013lna}. As the constraint from the Sun is superseded by others,~\eqref{eq:SunCriterion} suffices for our purposes.

\begin{figure}[t]
\begin{center}
\includegraphics[width=0.5\textwidth]{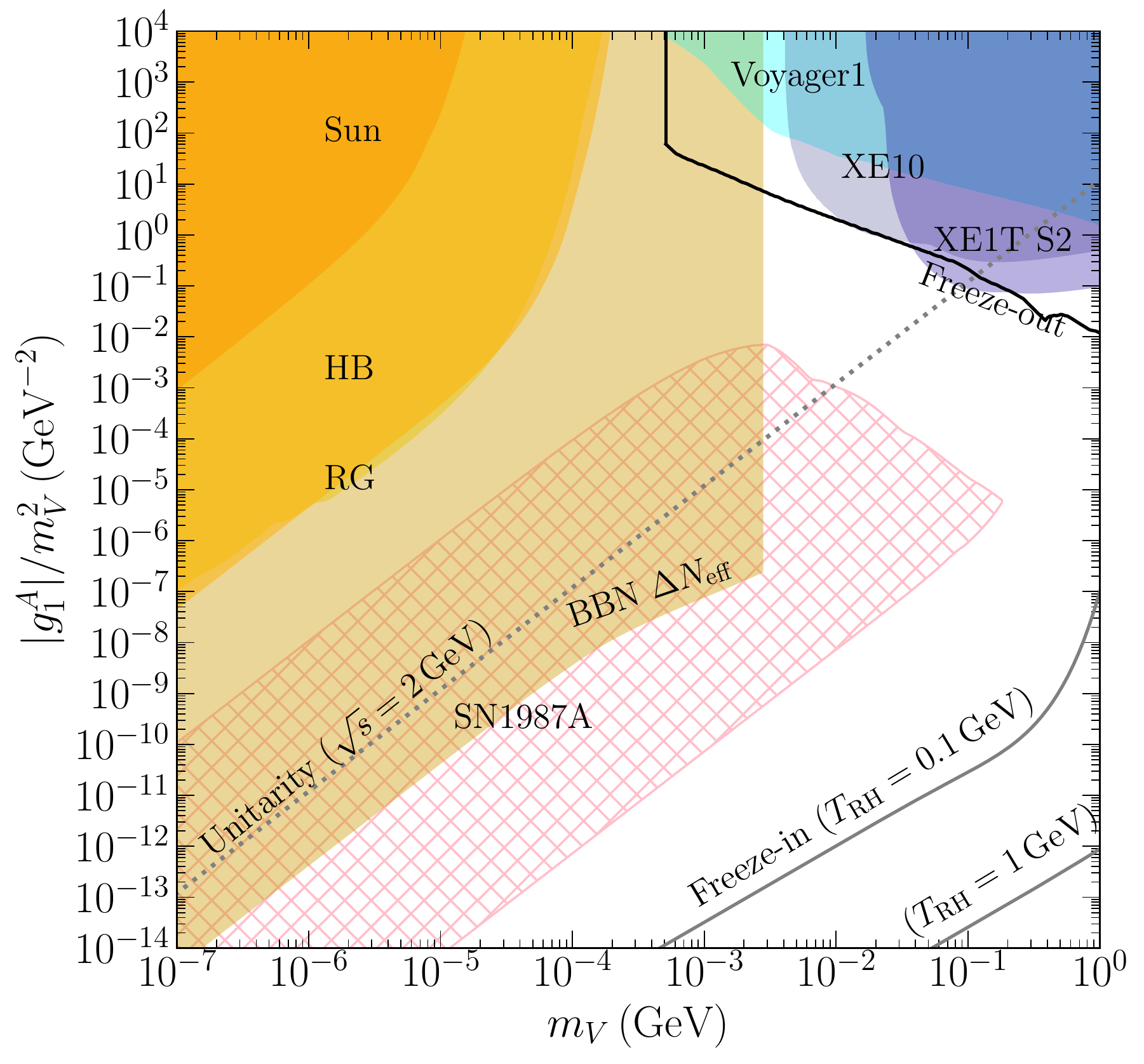}
\end{center}
\caption{Constraints on $|g_1^A|/m^2_V$, or, equivalently, on $|g_1^\Lambda|/\Lambda^2$, as a function of vector mass.  Stellar energy loss bounds from the Sun, HB and RG are effective for sub-keV $m_V$ (shaded regions), while SN1987A can probe $m_V$ up to 200\,MeV (hatched region). The solid  lines show the values for thermal freeze-out and  freeze-in, with reheating temperatures of 0.1\,GeV and 1\,GeV as lableled, to yield the observed DM abundance. The limits are in part superseded by the BBN constraint $\Delta N_{\rm eff}$ constraint. 
For $m_V\gtrsim 1\,\MeV$ additional constraints from DM-electron scattering in the direct detection experiments XENON10 and XENON1T and from indirect detection (Voyager~1) apply when assuming $V$ is DM. The dotted line provides the requirement of perturbative unitarity if no other new physics appears at $\sqrt{s}=2\,$GeV. }
\label{fig:stellar_bound1}
\end{figure}

To derive the energy loss rate of each environment, we consider the production of $V$ via plasmon decay, Compton-like scattering and electron-nucleus bremsstrahlung.
The analytical formulas for each respective process is given in our preceding work~\cite{Chu:2019rok}, which can be applied here by substituting $f(s)$ of Tab.~\ref{tab:results}.
In summary, the total energy loss rate for RG, HB stars and our Sun can be expressed as 
\begin{align}
   \dot{Q} = \dot{Q}_{\rm plasmon} + \dot{Q}_{\rm Compton} + \dot{Q}_{\rm brem}\,.
\end{align}
The inequalities above are then used to derive the upper bounds on the portal interactions for each dark state mass $m_V$, as long as its production in stellar objects is kinematically allowed.

Resulting constraints on the parameter space are shown in Figs.~\ref{fig:stellar_bound1}, \ref{fig:stellar_dim5} and \ref{fig:stellar_bound27}. A general trend to observe is that among the Sun, HB, and RG systems, RG stars yield the most stringent constraint. This is traced back to the fact that RG stars possess the highest core temperature as detailed in Tab.~\ref{tab:environment}. We are probing higher dimensional operators for which the production is UV-biased and the rates grow with available center-of-mass energy.

\subsection{SN1987A}

The MeV vector mass range is probed by the emission of $V$ from the  proto-neutron star (PNS) of SN1987A, assuming SN1987A was neutrino-driven supernova explosion. 
As a conservative criterion we require that the luminosity of $V$ shall not exceed the neutrino luminosity in the cooling phase,
\begin{align}
\label{eq:SN_bound}
   \int_{\rm core} dV\,\dot{Q}_{\rm SN} < L_\nu \simeq 3\times 10^{52}\,{\rm erg/s}\,,
\end{align}
where $L_\nu$ is taken at one second after the core bounce~\cite{1996slfp.book.....R}.
Since the positron abundance in the stellar core is not suppressed, 
the main $V$ pair-production mechanism is electron-positron annihilation~\cite{Chu:2019rok}.
To the latter we also add the contribution from plasmon decay.
In our numerical evaluation we take into account the  thermal masses of photons and electrons in the relevant production rates.
The total energy loss rate for PNS reads
\begin{align}
  \dot{Q} \simeq \dot{Q}_{\rm ann} + \dot{Q}_{\rm plasmon}\,.
\end{align}

In the low-coupling regime, $V$ streams freely after its production, escapes the star and~\eqref{eq:SN_bound} applies directly. 
On the other hand, once effective EM couplings are large enough, $V$ engages in a random walk with SM particles and can eventually be trapped inside the SN, rendering the energy loss argument ineffective.
For the latter, we follow the treatment in~\cite{Chu:2019rok} to derive the upper boundary of SN1987A constraint. Concretely, we first estimate the radius at which a thermalized  blackbody luminosity of $V$ equals the critical neutrino luminosity $L_\nu$, referred to as $r_d$. Taking the stellar model from~\cite{Fischer:2016cyd}, the value of $r_d$ varies from  29\,km for $m_{V}\lesssim 1$\,MeV to 11\,km for  $m_{V}\sim 400$\,MeV.  We consider $V$ as being sufficiently trapped once 
\begin{equation}\label{eq:SNupper}
    \int^{r_{\rm PNS}}_{r_d} dr \sum_{N=p,n} {\rho_N \over m_N} \sigma^{\rm VN}_{\rm T} \lesssim 2\, 
\end{equation}
is satisfied, where  $r_{\rm PNS} = 35\,$km is the PNS size, $\rho_N$ is the nucleon energy density and $m_N$ is the nucleon mass. For the evaluation we compute the momentum-transfer cross sections of~$V$ scattering on both, protons and neutrons, inside the PNS, defined by 
\begin{equation}\label{eq:sigmaT}
    \sigma^{\rm VN}_{\rm T} = \int^{1}_{-1} d\cos\theta (1-\cos\theta){d\sigma^{\rm VN} \over d\cos\theta } \,,
\end{equation} 
where $\theta$ is the scattering angle in the CM frame. Note that the dominant contribution always comes from $V$ scattering off protons via the proton's charge. In the evaluation, we include a form factor for the proton, adopting a dipole-form~\cite{Perdrisat:2006hj},
\begin{equation}
    F_E^p(t) = {1 \over (1+{|t|/\text{GeV}^2 \over 0.71} )^2}\,\text{,~and~}F_M^p(t) \simeq 2.79 F_E^p(t) \,. 
\end{equation}
Moreover, as $m_N$ is much larger than the temperature in the PNS, we assume that nucleons are at rest. In the end, Eq.~\eqref{eq:SNupper} gives the upper boundaries of our exclusion region from SN1987A.

\begingroup
\renewcommand{\arraystretch}{2.5} %
\begin{table*}[t]
    \centering
    \begin{tabular}{r|c@{\hskip 10pt}c@{\hskip 10pt}c@{\hskip 10pt}c}
    \toprule
       interaction type & $f(s)$ & $A_{ll}$ & $A_{\gamma \gamma}$ & $\overline{|\mathcal M_{Ve}(q)|}^2$  \\
\midrule
 magnetic dipole & $\dfrac{\mu_V^2 s(s-4m_V^2)(16m_V^2 +3s)}{12m_V^2}$ & $ \dfrac{7}{54} \mu_V^2 m_V^2 v^2 $ & $\dfrac{7 \mu_V^4 m_V^4 }{9\pi }$ & $\dfrac{64}{3} \pi \alpha  \mu_V^2 m_V^2$ \\
 electric dipole & $ \dfrac{d_V^2 s(s-4m_V^2)^2}{6m_V^2}$ & $ \dfrac{1}{108} d_V^2 m_V^2 v^4 $ & $\dfrac{d_V^4 m_V^4 v^2}{432 \pi }$ & $ \dfrac{256 \pi  \alpha  d_V^2 m_e^2 m_V^2}{3 q^2} $ \\
 electric quadrupole & $\dfrac{Q_V^2 s^2 (s-4m_V^2)}{16}$ & $\dfrac{1}{72} m_V^4 Q_V^2 v^2$ & $\dfrac{m_V^8 Q_V^4 v^2}{3456 \pi }$ &
$ 8 \pi  \alpha  m_e^2 m_V^2 Q_V^2 $ \\
 magnetic quadrupole & $\dfrac{\tilde{Q}_V^2 s^2 (s+8m_V^2)}{24}$ & $\dfrac{m_V^4 \tilde Q_V^2}{9}$ & $\dfrac{13 m_V^8 \tilde Q_V^4}{288 \pi }$ & $  \dfrac{4}{3} \pi  \alpha  m_V^2 q^2 \tilde Q_V^2  $ \\
 charge radius & $\dfrac{e^2 (g_1^A)^2 s^2 (s-4m_V^2) (12m_V^4 - 4m_V^2 s+ s^2)}{48 m_V^8}$ & $\dfrac{2}{9} \pi  \alpha  (g_1^A)^2 v^2 $ & 0 & $ \dfrac{128 \pi^2  \alpha^2  (g_1^A)^2  m_e^2 }{m_V^2} $\\
 toroidal moment & $\dfrac{e^2 (g_4^A)^2 s^3 (s-4m_V^2)}{3 m_V^6}$ & $\dfrac{32}{27} \pi  \alpha   (g_4^A)^2 v^2  $ & 0 
 & $ \dfrac{64 \pi^2  \alpha^2 (g_4^A)^2  q^4}{3 m_V^4}$
 \\
 anapole moment &  $ \dfrac{e^2 (g_5^A)^2 s^2 (s-4m_V^2)^2}{3m_V^6} $ & $ \dfrac{8}{27} \pi  \alpha (g_5^A)^2 v^4 $ & 0 & $\dfrac{256 \pi^2  \alpha^2 (g_5^A)^2   q^2}{3 m_V^2}$ \\ 
          \bottomrule
    \end{tabular}
    \caption{Summary of results that feed into the computation of limits and relic density. The first column shows the phase-space integrated expression $f(s)$, with mass dimension-2,  for $V^\dag V$ production defined in~\eqref{eq:fofs}. The coefficients $A_{ll}$ and $A_{\gamma \gamma}$ multiply the annihilation cross sections~\eqref{eq:sigmaAnnll} and~\eqref{eq:sigmaAnngg} into charged leptons and photons, respectively. The last column lists the leading terms of the squared matrix elements $\overline{|\mathcal M_{Ve}(q =\alpha m_e )|}^2$ of  DM-electron scattering for $m_V\gtrsim 3 m_e$. }
    \label{tab:results}
\end{table*}
\endgroup

\section{\boldmath$V$ as dark matter}
\label{sec:VDM}

\subsection{Freeze-out}

The freeze-out of $V$-particles that have come into thermal equilibrium with SM is governed by the  $2 \rightarrow 2$ annihilation cross sections into fermion and photon pairs. In the non-relativistic velocity expansion the annihilation into charged leptons $l$ of mass $m_l$ is given by%
\footnote{In the numerical evaluation we use the fully relativistic total invariant cross section and compute thermal average and freeze-out following~\cite{Gondolo:1990dk}.}
\begin{align}
\label{eq:sigmaAnnll}
\sigma_{V^\dag V \rightarrow l^{+} l^{-}} v = A_{ll} \frac{\alpha}{m_V^2}\left(1+\frac{m_l^2}{2 m_V^2}\right) \sqrt{1-\frac{m_l^2}{m_V^2}}\,.
\end{align}
The coefficients $A_{ll}$ the various EM form factors are listed in Tab.~\ref{tab:results}.

Note that only the magnetic quadrupole moment is $s$-wave whereas all other cross sections are $p$- or $d$-wave in their velocity suppression, rendering indirect detection constraints comparatively less important.
The annihilation into hadronic final states below the QCD phase-transition can be estimated via 
\begin{align*}
\sigma_{V^\dag V \rightarrow \text { had }}(s)=\sigma_{V^\dag V\rightarrow \mu^{+} \mu^{-}}(s) \times R(\sqrt{s})\,,
\end{align*}
where for the experimentally measured $R$-ratio
we use the tabulated data from \cite{ParticleDataGroup:2022pth}. Finally, the annihilation cross section into photon-pairs can be written as
\begin{align}
\label{eq:sigmaAnngg}
\sigma_{V^\dag V \rightarrow \gamma \gamma} v= A_{\gamma \gamma} / m_V^2 \,.
\end{align}
This cross section is only non-vanishing for form factors that are not proportional to $k^2$ as they are otherwise identically zero for on-shell photons. The coefficients $A_{\gamma \gamma}$ are listed in Tab.~\ref{tab:results}.
We find that the cross sections are either $s$- or $p$-wave in their velocity dependence.

\begin{figure*}[t]
\begin{center}
\includegraphics[width=0.48\textwidth]{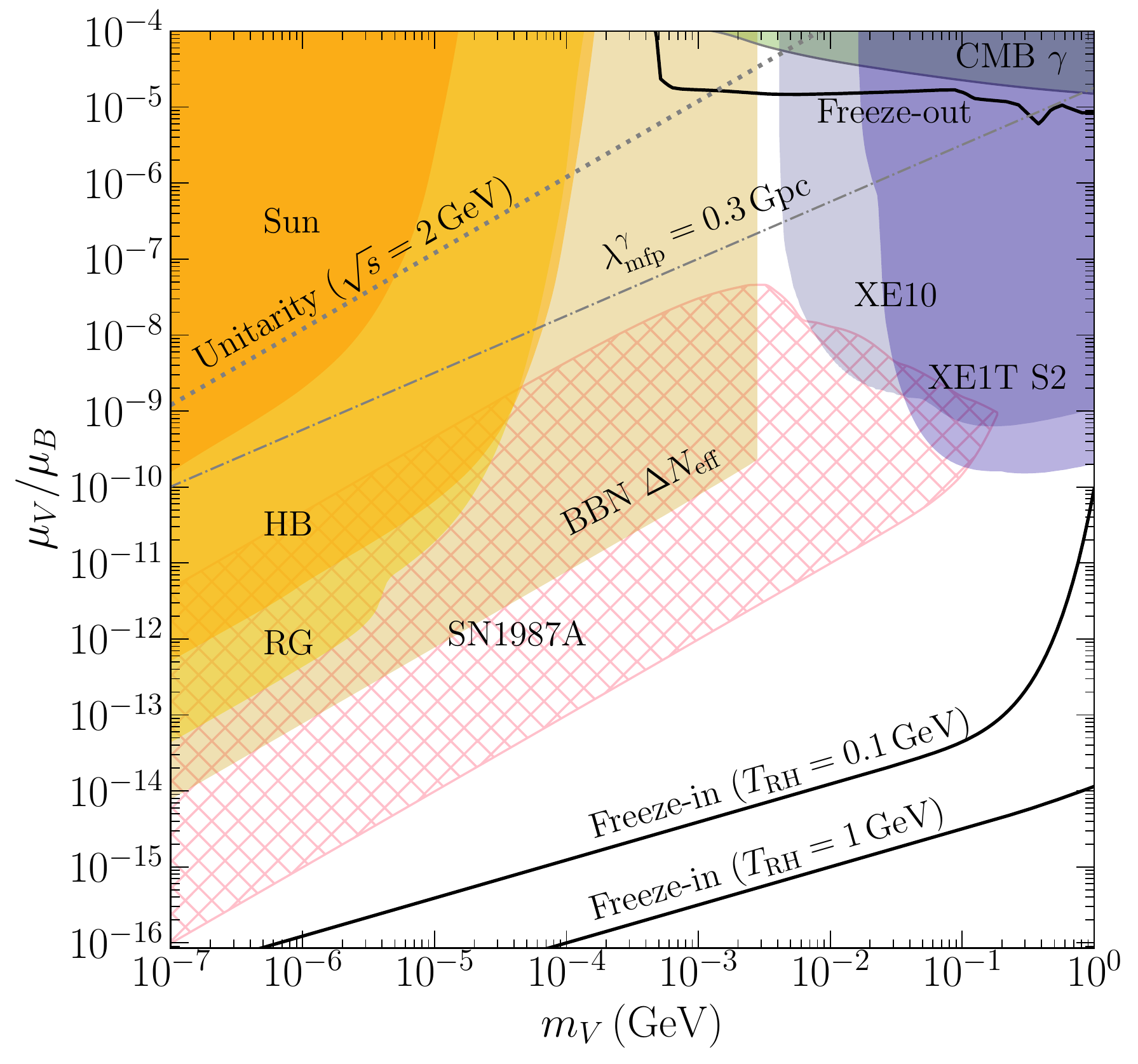}
~~\includegraphics[width=0.48\textwidth]{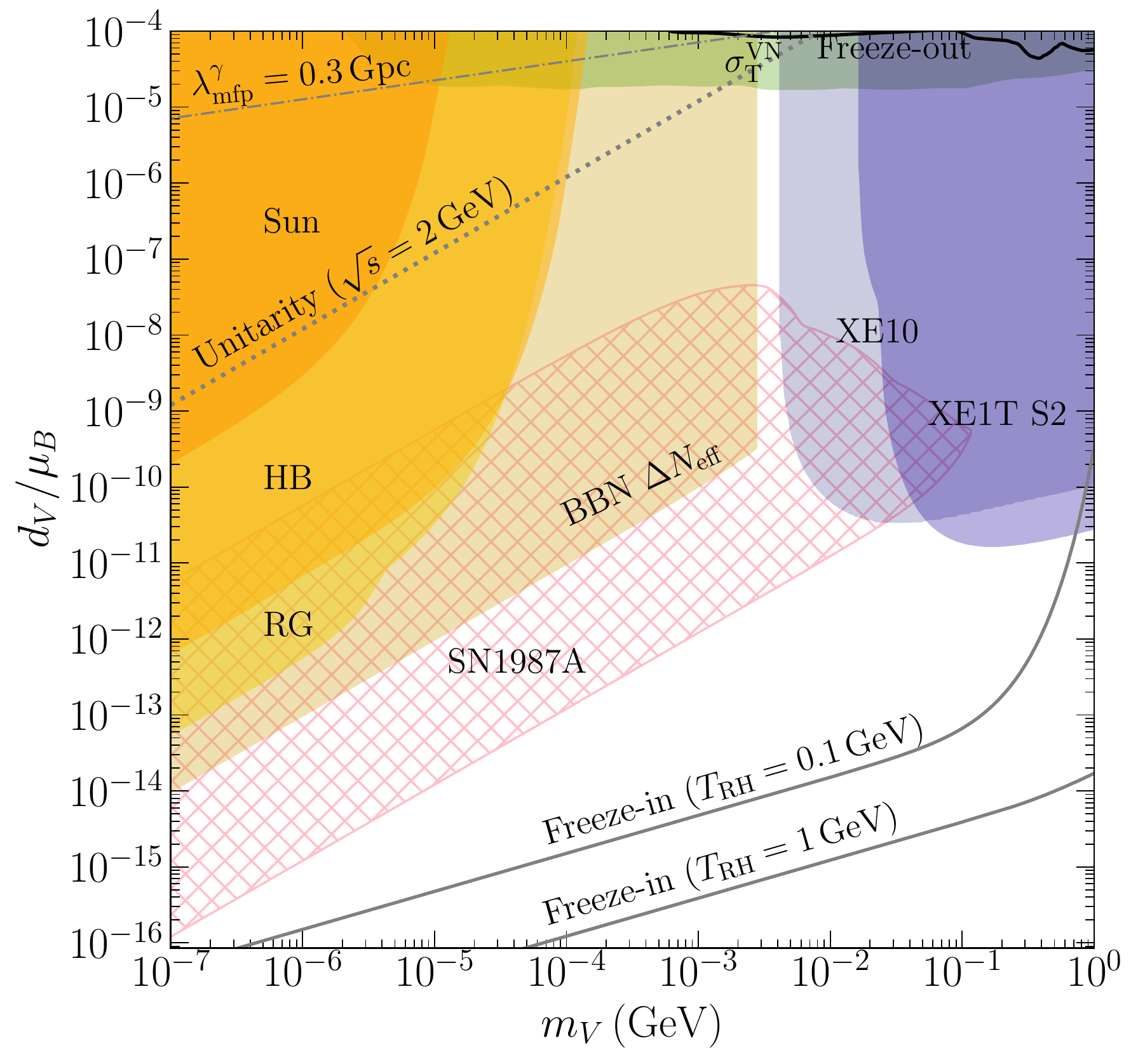}
\end{center}
\caption{Similar to Fig.~\ref{fig:stellar_bound1}, a summary of constraints derived in this work on magnetic and electric dipole form factors. Additional constraints that appear here are from gamma-ray transparency labeled $\lambda_{\rm mfp}^\gamma$, on energy injection during CMB and from scattering with protons, labeled by $\sigma_T^{\rm VN}$.}
\label{fig:stellar_dim5}
\end{figure*}
\subsection{Freeze-in}

We now consider the possibility that the production rate of~$V$ in the early Universe was always smaller than the Hubble rate, and $V$ never came into equilibrium with the SM. The freeze-in mediated by the higher dimensional effective operators considered in this work is UV dominated. The details then depend on whether the symmetry breaking of the UV theory happened before or after reheating.  For simplicity, here we take the example of $T_\text{RH} = 1$\,GeV and $0.1$\,GeV,  while assuming that all other particles in the UV theory are much heavier.  Using dimensional analysis, the results can then be easily re-scaled for other reheating temperatures as long as $T_\text{RH}\gg m_\text{V}$.%
\footnote{
For the freeze-in calculation of vector DM carrying magnetic dipole but without $Z_2$ custodial symmetry, see \cite{Krnjaic:2022wor}; freeze-in from magnetic and electric dipole moments of fermions was considered in~\cite{Chang:2019xva}.}

For the freeze-in production from the SM thermal bath, we solve numerically the Boltzmann equation
\begin{equation}\label{eq:Boltz}
 \dfrac{d n_V}{dt} +  3H n_V   \simeq  \sum_f n_f^2 \langle  \sigma_{_{\bar f f \to V^+V}} v\rangle +  n_W^2 \langle   \sigma_{_{W^+ W^-\to V^+V}} v\rangle \,,
\end{equation}
where $f$ denotes all SM charged fermions, and the sum takes into account the spin and color d.o.f. of these fermions. For the production cross sections, there are
\begin{equation}
    \sigma_{\bar f f \to V^+V} v   = \frac{\alpha  (2 m_f^2+s) }{2 s^3} f(s) \,, 
\end{equation}
and 
\begin{equation}
    \sigma_{ W^+ W^- \to V^+V} v  = \frac{\alpha  (s^2 + 20 m_W^2 + 12m_W^2)(s-4 m_W^2)}{72 m_W^4 s^3}  f(s) \,,%
\end{equation}
after electroweak symmetry breaking. To obtain the final DM abundance, a sudden thermalization after inflation is assumed, while the initial DM abundance is set to be zero. We also estimate that for reheating temperatures higher than the electron mass, the contribution from plasmon decay is very subleading, and thus neglected here.

It is worth emphasizing that the freeze-in  mechanism produces DM particles that in average carry kinetic energy close to the photon temperature. So,  while here the results are shown for the whole mass range,  DM with a mass below keV is excluded observationally,  being too hot to satisfy the Lyman-$\alpha$ constraints~\cite{Irsic:2017ixq,Villasenor:2022aiy}. Apparently, this exclusion also applies to the thermal freeze-out mechanism above. 

\subsection{\boldmath$N_{\rm eff}$ constraint from BBN}

Following the calculation above, we  also obtain bounds from Big Bang nucleosynthesis (BBN) by requiring the energy deposited in the dark sector should not exceed the BBN constraint on the extra relativistic degrees of freedom, $\Delta N_{\rm eff} \lesssim 0.407$~\cite{Yeh:2022heq} with the minimally-allowed reheating temperature $T_{\rm RH} \sim 4$\,MeV~\cite{Kawasaki:2000en, Hannestad:2004px}. That is, now we replace the number density  in Eq.~\eqref{eq:Boltz} with energy density of the dark sector as follows:
\begin{equation}
  \dfrac{d \rho_V}{dt} +  3H(\rho_V + p_V)     \simeq  n_e^2  \,\langle  \sigma_{{\bar e e \to V^\dag V}} v \cdot \sqrt{s} \rangle \,,
\end{equation}
and calculate the $\rho_V$ evolution from negligible initial value at $T_{\rm RH}$ to $T = 1\,$MeV,  where, because of the low temperatures involved, we only need to include the production from electrons (with an $O(1)$ correction from $\gamma\gamma\to V^\dag V$ for a subset of operators). Similar bounds are obtained by requiring electrons and $V$ are not mutually thermalized,~i.e., $n_e \langle \sigma_{e^+ e^- \to V^+V} v \rangle \le H(T)$ at $T\simeq 1\,$MeV. We cut off the bound at $m_V = 2.8\,$MeV, since a thermalized vector species with larger mass cannot yield $\Delta N_{\rm eff} \gtrsim 0.407$ at $T \simeq  1\,$MeV. We emphasize that these considerations guarantee that $V$ particles do not over-populate to jeopardize the standard BBN predictions. It is based on the assumption that $V$ particles behave either as dark radiation or as non-relativistic matter, depending on their average kinetic energy, and remain present hundreds of seconds after being produced. If $V$ particles decay sufficiently fast (see {\it e.g.} \cite{Arina:2009uq}), the BBN bounds may be alleviated. This class of scenarios may lead to novel signatures, depending on the specific decay channels. 

\subsection{Direct detection constraints}

The MeV mass region of EM interacting DM candidates is chiefly probed by the scattering on atomic or valence electrons  in direct detection experiments~\cite{Essig:2011nj,Essig:2012yx}.
Limits are often expressed in terms of a DM-electron reference cross section on free electrons where the squared matrix element is evaluated at a typical atomic squared momentum transfer
$q^2= \alpha^2 m_e^2 $~\cite{Essig:2011nj},
\begin{align}
\label{eq:sigmaelectron}
  \bar \sigma_e \equiv \frac{1 }{ 16\pi (m_e + m_V)^2 }
\overline{|\mathcal M_{Ve}(q =\alpha m_e )|}^2 \, .
\end{align}
We list the expressions for $\overline{|\mathcal M_{V e}(q)|}^2$ in Tab.~\ref{tab:results}.

The recoil cross-section for DM-electron scattering from atomic orbital $n,l$ is given by~\cite{Essig:2011nj},
\begin{align}
  \frac{  d \langle\sigma_{n,l} v \rangle }{ d\ln E_{e}  }  & =  \frac{\bar\sigma_e}{8\mu_e^2} \int dq\, \left[ q  |F_{\rm DM}(q)|^2
              |f^{\rm ion}_{nl}(p_e, q)|^2                                                         \right.
  \nonumber   \\ & 
                   \left.    \qquad \qquad  \quad \times
                   \eta(v_{\rm min}(q,\Delta E_{n,l})) \right] \,,
\end{align}
where $\eta(v_{\rm min})$ is the  velocity average of the inverse
speed,
$\eta(v_{\rm min}) = \langle  \Theta(v-v_{\rm min})/v
\rangle_{f_{\rm det}} $
over the  distribution $f_{
\rm det}$ of relative velocity in the detector frame; the minimum velocity $v_{\rm min}$ to inflict an electron recoil energy $E_e$ is given by $v_{\rm min}(q,\Delta E_{n,l}) \simeq q/(2\mu_e) + \Delta E_{n,l}/{q} $ where 
 $\Delta E_{n,l} = E_e + |E_{n,l}| $ with $E_{n,l} $ being the ionization threshold of the $n,l$ orbital. The momentum transfer dependence of the cross section is shifted into a DM form factor $|F_{\rm DM}|^2 =
 \overline{|\mathcal M_{V e}(q)|}^2 / \overline{|\mathcal M_{V e}(q=\alpha m_e)|}^2  $; the electron ionization form factors $ |f^{\rm ion}_{nl}(p_e, q)|^2 $ are taken from~\cite{Essig:2019xkx}.

We derive constraints on the vector DM parameter space by utilizing the results from the XENON10 and XENON1T experiments~\cite{XENON10:2011prx, Essig:2012yx, XENON:2021qze}. The modeling of the formation of the ionization-only S2 signals in these liquid scintillator experiments as well as the limit setting procedure follows~\cite{An:2017ojc}. The limits can be significantly extended to lower masses once the solar-reflected component of DM is included~\cite{An:2017ojc,Emken:2021lgc,An:2021qdl}. The results from the semiconductor experiments SENSEI~\cite{SENSEI:2020dpa} and DAMIC-M~\cite{DAMIC-M:2023gxo}
will also improve the obtained limits at the low mass end. For DM masses below one GeV, the bounds from nucleon recoil events, e.g., obtained by CRESST-III~\cite{CRESST:2019jnq}, are relatively weaker~\cite{Hisano:2020qkq}, and thus not included here.

\begin{figure*}[t]
\begin{center}
\includegraphics[width=0.49\textwidth]{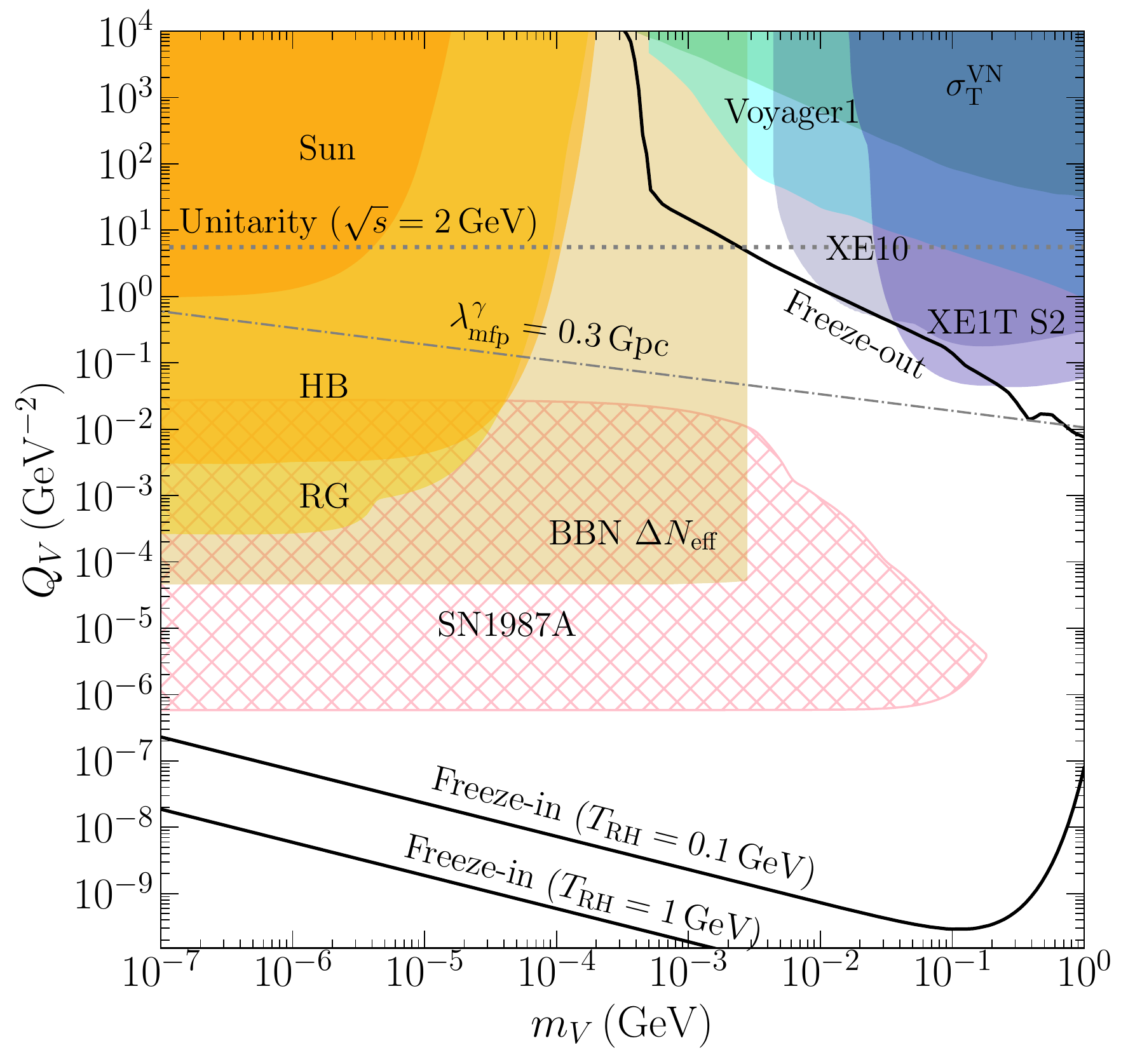}
\includegraphics[width=0.49\textwidth]{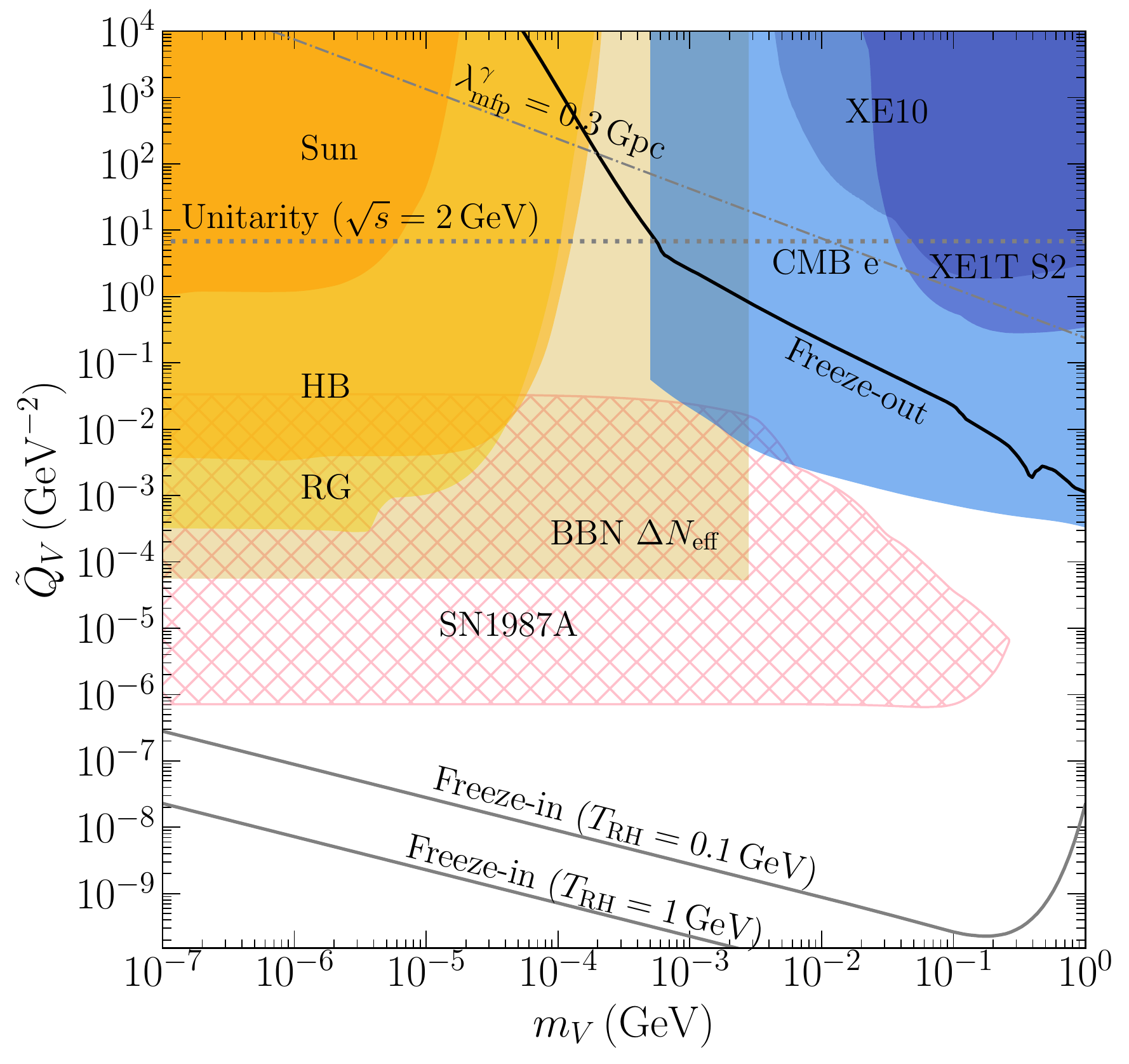}
\includegraphics[width=0.49\textwidth]{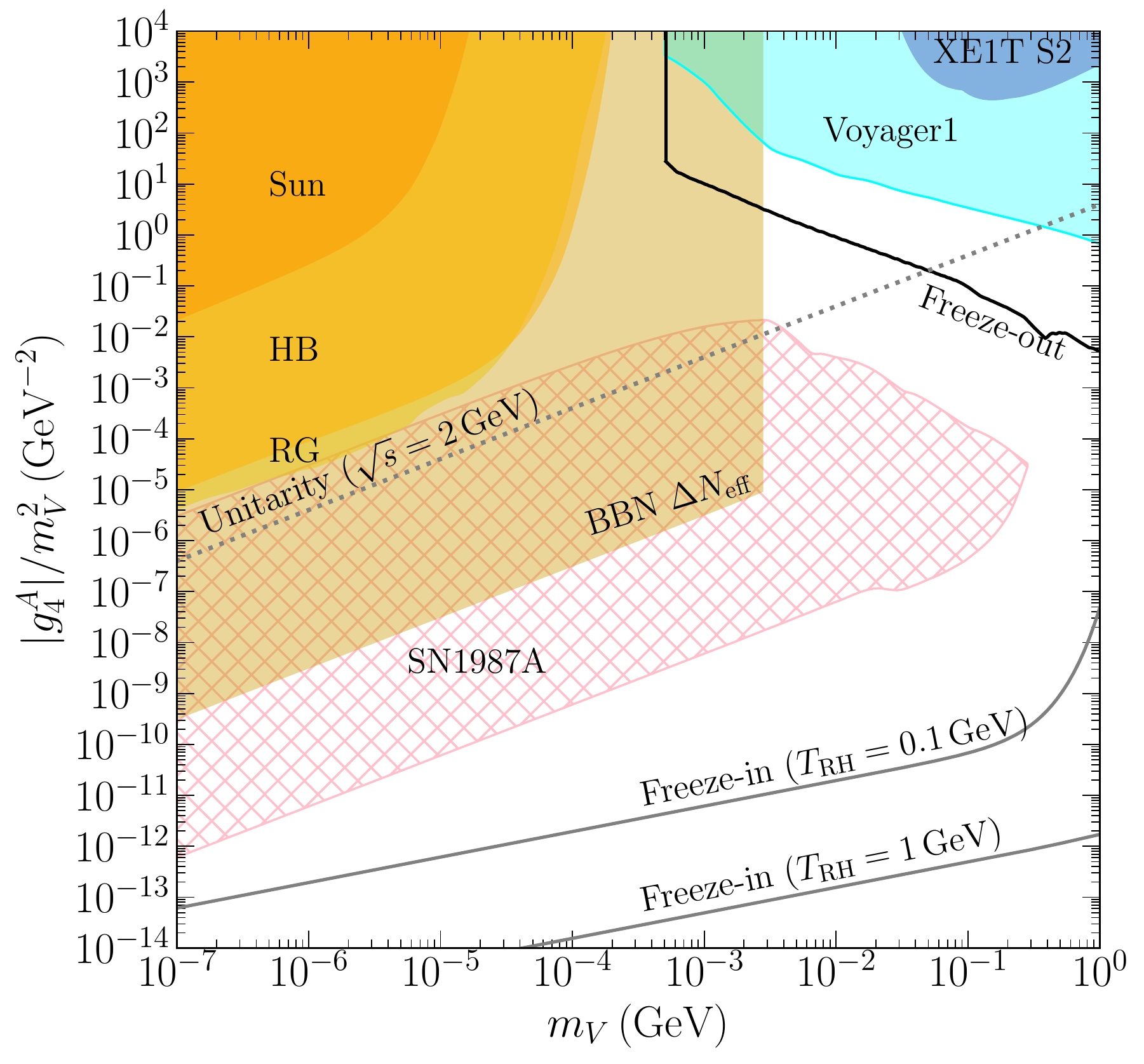}
\includegraphics[width=0.49\textwidth]{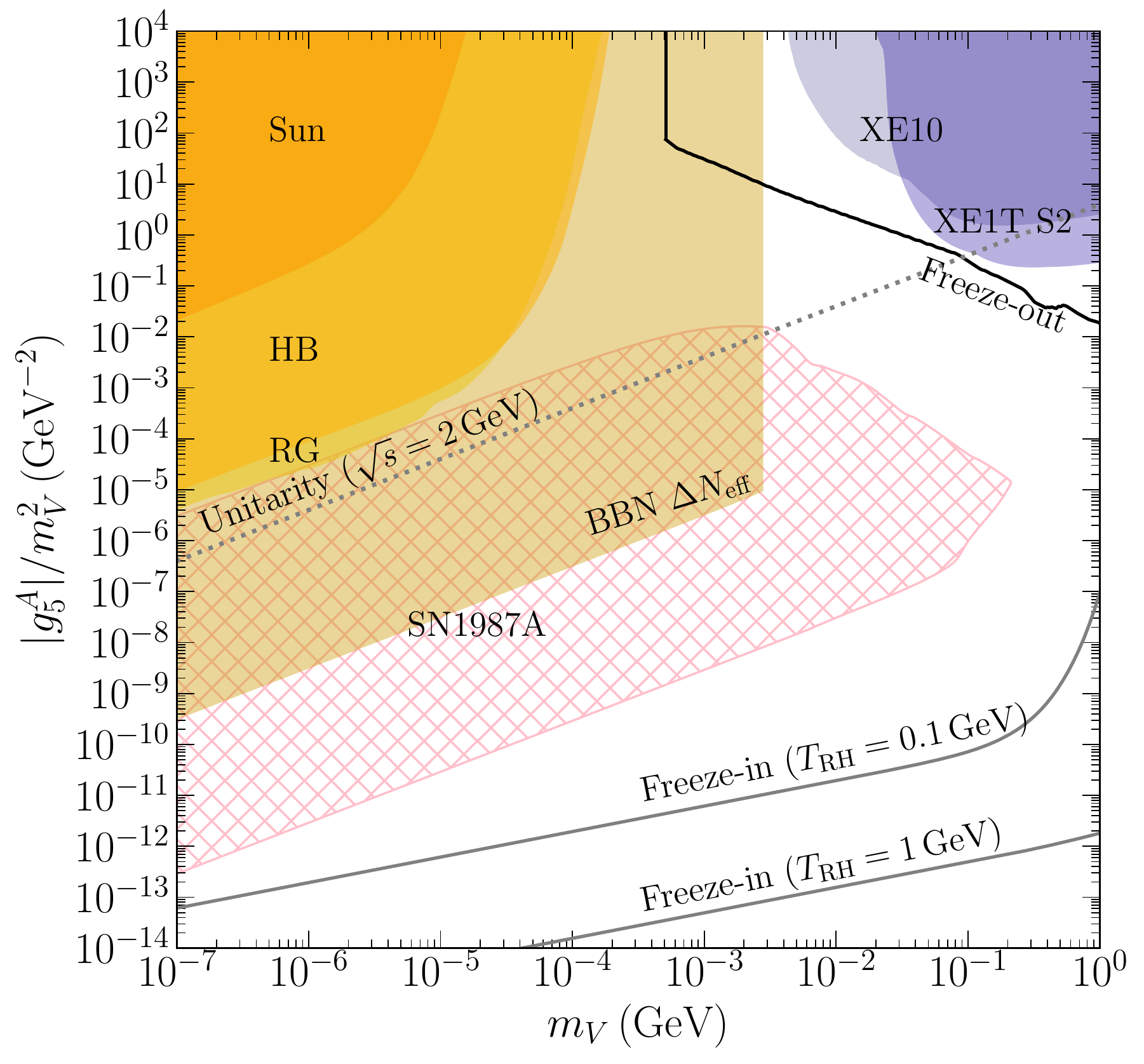}
\end{center}
\caption{Similar to Figs.~\ref{fig:stellar_bound1} and \ref{fig:stellar_dim5}, constraints on the  electric (magnetic) quadrupole moments $Q_V$ ($\tilde Q_V$) in the top panel and on the CP-violating toroidal moment ($g_4^A$) and CP-conserving anapole moment ($g_5^A$) in the bottom panel.}
\label{fig:stellar_bound27}
\end{figure*}

\subsection{Indirect search  of dark matter annihilation} 

Dark Matter annihilation into the visible sector is constrained by observables at low redshift, under the condition that this DM candidate is symmetric, and dominates the observed relic abundance.  For most of the operators studied here, the DM annihilation cross section is velocity-suppressed, and the corresponding limits are generally weak, allowing for the standard thermal freeze-out except for the magnetic quadrupole case. 

For the numerical results shown below,  we take Planck data~\cite{Planck:2018vyg} that  constrain both DM annihilation channels,  $VV\to \gamma\gamma$ and $VV\to e^+e^-$, at the epoch of cosmic microwave background (CMB) emission~\cite{Slatyer:2015jla} and during cosmic reionization~\cite{Liu:2016cnk}. 
For the limits on DM annihilating into two photons at present, we adopt the bounds that have been derived from several X-ray experiments (NuSTAR, INTEGRAL and COMPTEL)~\cite{Ng:2019gch, Laha:2020ivk},  as well as  gamma-ray observations from the EGRET and Fermi-Lat data~\cite{Boddy:2015efa, Fermi-LAT:2015kyq}. 
On the other side, the non-observation of a cosmic-ray excess puts upper limits on  DM annihilating into electron-positron pairs, among which the local $e^+e^-$  measurement by Voyager~1 provides the strongest constraints on our model~\cite{Boudaud:2016mos, Boudaud:2018oya}.  

For simplicity, only the most relevant bounds are shown in our constraint figures~\ref{fig:stellar_bound1}, \ref{fig:stellar_dim5} and~\ref{fig:stellar_bound27}. Moreover, alternative considerations that are able to probe DM annihilation with DM mass well below the MeV-scale, such as gas heating~\cite{Wadekar:2021qae}, are not included, as they are still much weaker than the stellar and BBN constraints at this moment.

\subsection{Cosmological limits on  DM-SM scattering} 

If $V$ is the dominant DM component and sufficiently scatters with  protons or electrons, it leads to modifications of the CMB spectrum, as well as the matter power spectrum; see~e.g., the recent works~\cite{Slatyer:2018aqg, Kumar:2018yhh, Cappiello:2018hsu, Maamari:2020aqz, Nguyen:2021cnb, Rogers:2021byl,Li:2022mdj}. Here we consider the limits from the DM-proton scattering, using the results in    \cite{Maamari:2020aqz, Nguyen:2021cnb}. To obtain the upper bounds on the coefficients, we use the momentum-transfer cross sections calculated from Eq.~\eqref{eq:sigmaT}, labeled as ``$\sigma_T^{\rm VN}$'' in our constraint figures. 
Since these limits are derived using observables inferred from epochs in the  Universe where DM is already extremely non-relativistic, they do not strongly constrain the effective operators studied here. 

In contrast, very high energy (VHE) photons scattering with the dark matter medium may result in much stronger limits. For instance, the attenuation of VHE $\gamma$-rays have been used to measure the density of extragalactic background light in space; see e.g.~\cite{HESS:2017vis}.  Nevertheless, the validity of our effective operator approach is not guaranteed in such high-energy collisions.  Here, we instead provide a benchmark line illustrating the parameters for which a photon with $E_\gamma =1$\,TeV has a mean-free-path of 0.3\,Gpc. This is comparable to the actual mean-free-path of a TeV photon propagating in the extragalactic background light. 

Similarly to the DM-proton scattering case above,  we use the momentum-transfer cross sections of DM-photon scattering, but now in the frame of the non-relativistic DM medium. Consequently, the mean-free-path of VHE photons in the DM medium can be expressed as
\begin{equation}
    \lambda^\gamma_{\rm mfp} = 0.3 \,\text{Gpc}  {m_V\over \rho_V \sigma_T^{V\gamma} } \simeq \left({m_V \over \text{MeV}}\right) \left({10^{-24}\text{cm}^2   \over  \sigma_T^{V\gamma}}\right) \,, 
\end{equation}
where we have taken the average DM density $\rho_V  \simeq 1.2\times 10^{-6}\,$GeV/cm$^3$~\cite{Planck:2018vyg}. Dimensional analysis suggests that $\sigma_T^{V\gamma}$ scales as $E_\gamma^5 \mu_V^4 /m_V^3$,  $E_\gamma^3 d_V^4 /m_V$, $E_\gamma^5 Q_V^4  m_V$ and $E_\gamma^3 \tilde Q_V^4 m_V^3$ for the first four operators and vanishes for $g^A_{1,4,5}$. The line labeled as ``$\lambda^\gamma_{\rm mfp} = 0.3 \,\text{Gpc}$'' in Figs.~\ref{fig:stellar_dim5} and ~\ref{fig:stellar_bound27} can be considered an upper limit for as long as the effective operator approach is valid at a center of mass energy of $\sqrt{m_V \text{TeV}}$ (see the discussion in Sec.~\ref{sec:UV}.) 
Note that stronger bounds may be obtained from considering Blazar photons with much higher energies, as well as the existence of a DM spike around the source~\cite{Ferrer:2022kei}.

\section{Validity of the effective description and the limit \boldmath$m_V\to 0$}
\label{sec:UV}

In this section, we address important questions on the validity-range of the effective operator description and on the limit of diminishing vector mass.

\subsection{Perturbative unitarity}

As is well known, the amplitudes involving on-shell massive vectors may contain factors of $s/m_V^2$ with $\sqrt{s}$ being the C.o.M.~energy, and thus lead to bad high-energy behavior. Here we focus on the elastic scattering process $V^\dag V\to V^\dag V$, and require its cross section to be below the unitarity limit~\cite{Dicus:1973gbw, Lee:1977yc, Lee:1977eg}  as follows: 
\begin{equation}
    \sigma_{V^+V\to V^+V}(s) \lesssim {4\pi \over s}\sum_l (2l+1)\,,
\end{equation}
where $l$ stands for the contribution of $l$-partial wave scattering. We estimate the corresponding limit for each effective operator, by only including the s-channel process via an intermediate photon, where one partial wave dominates the cross section.\footnote{A rigorous derivation should include both s/t-channel processes and separate each partial wave contribution to scattering amplitude, \emph{e.g.}, $i{\mathcal M}_{{V_L^+V_L\to V_L^+V_L}}$, using vector polarization tensors.  } This inequality needs to be satisfied for the values of $\sqrt{s}$ for the processes studied in this paper. We checked this and affirm that our derived exclusion bounds are indeed self-consistent. The inelastic process of $V$-creation,  $\bar f f \to V^\dag V $, automatically satisfies the unitarity limit, as it is further suppressed by the EM fine-structure constant~$\alpha$. 

We may, however, go further and ask: given $m_V$ and a value of C.o.M.~energy $\sqrt{s}$, what is the maximally allowed value of the effective coupling constant, below which perturbative unitarity remains respected?  As an example, we choose $\sqrt{s} = 2\,$GeV and show the corresponding coupling constant values as dotted gray lines in Figs.~\ref{fig:stellar_bound1},~\ref{fig:stellar_dim5} and~\ref{fig:stellar_bound27}. That is to say, in the region above those lines, a dark Higgs particle must enter the theory at or below the considered energy scale to restore unitarity. As we exclusively consider processes with $\sqrt{s}\lesssim 2\ \GeV = 2 m_V|_{\rm max}$, the bounds derived in this work are valid above the dotted gray lines, but if other higher energy probes such as collider constraints are considered, one should not rely on the effective operator picture presented above.

\subsection{An exemplary UV-model}

The study presented above in terms of effective multipole couplings of $V$ to the photon stands by itself, but as is pertinent to the physics of massive vector bosons, the limit $m_V \to 0$ deserves special attention. Indeed, in Figs.~\ref{fig:stellar_bound1},~\ref{fig:stellar_dim5} and~\ref{fig:stellar_bound27} we observe a strengthening of stellar bounds as the vector mass diminishes. This, of course, does {\it not} mean that the production rate diverges as $m_V\to 0$. To see this, however, one must make reference to a UV description that gives rise to the effective operators in~\eqref{eq:effectiveL}. 

A UV model that gives rise to all effective couplings in~\eqref{eq:effectiveL} except $g_4^A$ was presented by some of us in~\cite{Hisano:2020qkq}. Here, we briefly outline the main ingredients;  a detailed description is found in the original work. Under a dark SU(2)$_D$ gauge symmetry, a vector triplet $W_D^a$, a dark Higgs doublet $\Phi_D$ as well as a fermion doublet $\Psi_l$ and singlet $\Psi_e$ are introduced. Spontaneous symmetry breaking by the vacuum expectation value $\langle \Phi_D\rangle = v_D/\sqrt{2}$ yields a common vector boson mass $m_{W_D} = g_D v_D/2$, where $g_D$ is the  SU(2)$_D$ gauge coupling. The masses of fermions receive additional Yukawa contributions by the breaking. In a mass-diagonal basis one is left with massive fermions $\Psi_N $ and $\Psi_E^{1,2}$ with respective masses $m_N$ and $m_{E^i}$ and electric charges $-e$ that originate from a non-trivial hypercharge assignment of $\Psi_l$ and $\Psi_e$. The interaction Lagrangian then reads~\cite{Hisano:2020qkq},
\begin{align}
\label{eq:LintUV}
\mathcal{L}_{\mathrm{int}}= &-\frac{g_D}{\sqrt{2}}\left(\bar{\Psi}_E^i\left[\left(V_L\right)_{1 i} P_L+\left(V_R\right)_{1 i} P_R\right] \gamma^\mu \Psi_N W_{D \mu}^{-}+\text {h.c.}\right) \nonumber \\  & -e \Psi_N \gamma^\mu \Psi_N A_\mu-e \bar{\Psi}^i{ }_E \gamma^\mu \Psi_E^i A_\mu .
\end{align}
Here $V_{L,R}$ are the unitary matrices that diagonalize the fermions; $P_{L,R}$ are chiral projectors. Here, $W_{D}^{\pm}$ refers to the components of the vector triplet that carry a custodial (global) $U(1)_D$ symmetry ensuring their stability; $W_D^0$ is unprotected and together with  $\Psi_N $ and $\Psi_E^{1,2}$ decay to the SM; see~\cite{Hisano:2020qkq}. In the language of the previous sections we may  then assign $V= W^-$ and $V^\dag = W^+$ and identify $m_V = m_{W_D}$.  

\subsection{The limit \boldmath$m_V\to 0$}

The multipole moments of~\eqref{eq:effectiveL} are then  radiatively induced by the interactions in~\eqref{eq:LintUV} through triangle diagrams where the electrically charged states $\Psi_N $ and $\Psi_E^{1,2}$ run in the loop. The explicit expressions in the limit $m_V\ll m_N,\ m_{E^i}$ are given in App.~\ref{app:uvcouplings}. Here, we are principally interested in connecting the scalings of couplings and emission rates with the UV-parameters of the theory in the limit $m_V\ll T_{\rm star}\ll m_{N,E^i}$.

\begingroup
\renewcommand{\arraystretch}{2.5} 
\begin{table} 
 \centering  
  \begin{tabular}{l|cccc}
  \toprule 
  Coupl.~ & UV model  &  $\dot Q\propto f(s)$ &  $ \dot Q|_{m_{_V}\to 0}$ &  pol. \\
  \midrule 
  $\mu_V$   & ${ \dfrac{g^2_D}{m_V}} \propto \dfrac{g_D}{v_D}$    & ${\dfrac{\mu_V^2}{m^2_V}}\propto \dfrac{ 1}{{v}^4_D} $ & finite & all  \\
$Q_V$      & $\dfrac{ g^2_D}{{m_V^2}} \propto \dfrac{1}{ {v}_D^2 } $   & $Q_V^2 \propto \dfrac{ 1}{ {v}^4_D} $  & finite & LL,TT \\
 $g_1^A$    & $\dfrac{ g^2_D m_V^2 }{ {m_N^2}} \propto  \dfrac{g^4_D {v}_D^2}{ m_N^2 } $    & $\dfrac{(g_1^A)^2 }{ m_V^8} \propto \dfrac{ 1}{ {v}^4_D} $ &  finite & LL,TT\\
 \midrule
   $d_V$     & $\dfrac{ g^2_D}{{m_V}} \propto\dfrac{ g_D}{{v}_D } $   &  $\dfrac{d_V^2}{m^2_V} \propto \dfrac{ 1}{{v}^4_D}  $ &  finite  & TT\\
 $\tilde Q_V $ & $\dfrac{ g^2_D}{ {m^2_V}} \propto\dfrac{ 1 }{ {v}^2_D} $      & $\tilde Q_V^2 \propto \dfrac{ 1}{ {v}^4_D} $ & finite & LT, TT\\
 \midrule
   $g_4^A$   &  0  & $\dfrac{(g_4^A)^2 }{ m_V^6} $  & --  & LT\\
   \midrule
   $g_5^A $ & $\dfrac{ g^2_D m_V^2 }{ {m_N^2}} \propto  \dfrac{g^4_D {v}_D^2}{ m_N^2 } $  & $\dfrac{(g_5^A)^2 }{ m_V^6} \propto   \dfrac{ g^2_D}{ {v}^2_D}  $   & 0& LT \\
  \bottomrule
 \end{tabular} 
 \caption{Scaling of operators and stellar emission rates. The first column gives  the EM moments in terms of UV parameters (dark gauge coupling $g_D$ and symmetry breaking scale $v_D$). The middle column shows the scaling of the stellar energy loss rates $\dot Q$. 
The last column shows the behavior of $V$-production rates in the $m_V\to 0$ limit under the condition that ${v}_D \gg T_{\rm star}$ so that the dark Higgs remains decoupled.}
\label{tab:scaling}
\end{table}
\endgroup

\begingroup
\renewcommand{\arraystretch}{2.5} %
\begin{table}  
 \centering  
  \begin{tabular}{l|c|c|c|c|c|c|c}
  \toprule 
 & $\kappa_\Lambda$  & $\lambda_\Lambda$  & $g_1^A$ & $ \tilde \kappa_\Lambda$ & $\tilde \lambda_\Lambda $ & $ g_4^A$  & $g_5^A $\\
 \midrule 
 UV & $g_D^2$ & $\dfrac{g_D^2 \LambdaUV^2}{m_N^2}$ &  $ \dfrac{g_D^2 m_V^2}{m_N^2}$ & $g_D^2$ & $ 0 $ & 0 & $\dfrac{g_D^2 m_V^2}{m_N^2}$ \\[-4ex]
 &   \multicolumn{3}{c|}{$\underbrace{\hphantom{\qquad\qquad\qquad\qquad\quad}}$} &   \multicolumn{2}{c|}{$\underbrace{\hphantom{\quad\qquad\qquad}}$} & \\[-1ex]
 $C,\!P$ &  \multicolumn{3}{c|}{$(+,+)$}   & \multicolumn{2}{c|}{$(+,-)$}  & $(-,+)$ & $(-,-)$  \\ 
 $\dot Q_{\rm LL}$  &  \multicolumn{3}{c|}{$ \dfrac{\kappa_\Lambda^2}{m_V^4} \propto \dfrac{g_D^4}{m_V^4} $} &  \multicolumn{2}{c|}{0} & 0 & 0 \\
  $\dot Q_{\rm LT}$ &  \multicolumn{3}{c|}{$ \dfrac{\kappa_\Lambda^2}{m_V^2} \propto \dfrac{g_D^4}{m_V^2} $} & \multicolumn{2}{c|}{$ \dfrac{\tilde\kappa_\Lambda^2}{m_V^2} \propto \dfrac{g_D^4}{m_V^2} $} & -- &  $\dfrac{(g_5^A)^2 }{ m_V^6} \propto \dfrac{g_D^4}{m_V^2} $   \\
 $\dot Q_{\rm TT}$ &  \multicolumn{3}{c|}{$\left(\dfrac{\lambda_\Lambda}{\LambdaUV^2}+ \dfrac{g_1^A}{m_V^2}\right)^2\propto  {g_D^4}$ } &  \multicolumn{2}{c|}{$\tilde \kappa_\Lambda^2 \propto g_D^4$} & 0 & 0 \\ 
  \bottomrule
 \end{tabular} 
 \caption{Operators grouped by their $C$ and $P$ transformation properties. The first row shows the leading scaling of the operator when $m_V$ is the smallest scale in the problem, the second row details their discrete symmetries, and the subsequent rows show the scaling of the stellar emission rate $\dot Q_{\lambda\lambda'}$ in their combination
 with final state vector polarity $\lambda, \lambda'$ in the limit $m_V\to 0$. 
As can be seen, the various strengths of emission are commensurate with the expectation in~\eqref{eq:polscaling}. 
 }
 \label{tab:scalingUV}
\end{table}
\endgroup

On general grounds, from the UV perspective, one expects the following scaling of emission rates in the high energy limit $\sqrt s/m_V \gg 1 $ for the various combinations of vector boson polarities,
\begin{align}
\label{eq:polscaling}
    \dot Q_{\lambda\lambda'} \propto \begin{cases}
    g_D^4/ m_V^4  & \lambda \lambda'={\rm LL} , \\
    g_D^4 / m_V^2 & \lambda \lambda'={\rm LT} , \\
    g_D^4  & \lambda \lambda'={\rm TT} . 
    \end{cases}
\end{align}
The differences are of course traced back to the relative enhancement of longitudinal (L) over transverse (T) modes for which the respective polarization vectors read
\begin{equation}
    \epsilon_{\rm L} = \left({p\over m_V}, 0, 0, {E\over m_V}\right)\,,~\epsilon^{\pm}_{\rm T} = \left(0, {1 \over \sqrt{2}}, \pm{ i\over \sqrt{2}}, 0\right)\,.
\end{equation}
For example, in the UV picture, the emission rate for $V_L V^\dag_L$ is  proportional to   
\begin{equation}\label{eq:LLscaling}
   \dot  Q_{\rm LL} \propto |(g_D \epsilon_{L,1}) (g_D\epsilon_{L,2})|^2 \propto {g_D^4 \over m_V^4} \propto { 1 \over {v}^4_D} \, ,
\end{equation}
independent of $g_D$. This is because the L mode of $V$ is equivalent to the emission of Goldstone bosons that attach to the triangle graphs that induce the effective coupling with Yukawa strength. 

We summarize the results obtained in terms of the effective couplings in Tab.~\ref{tab:scaling}. The first column shows the leading scaling of the coupling. The second column shows the corresponding scaling of the stellar emission rates in the limit $m_V\ll T_{\rm star}$, the penultimate column highlights the corresponding limit and the last column shows the available final polarization states available through the operator. As can be seen, the rates are manifestly finite 
 and there is no divergence at $g_D\to 0$.%
 \footnote{
Note that all fermion masses and ${v}_D$ are assumed to be much larger than the stellar temperature. If, instead, one allows ${v}_D$ to be lowered, the contribution of the dark Higgs will likely need to be taken into account in order to regularize the production rates, in analogy to the role of the SM Higgs in high-energy $W_L W_L$ scattering.}

As can be seen from  Tab.~\ref{tab:scaling} the scaling~\eqref{eq:LLscaling} is indeed observed for $Q_V$ and $g_1^A$ which permit the LL mode.
 However, Tab.~\ref{tab:scaling} also reveals that effective operators that do not permit the LL mode in the final state, show the same scaling as in~\eqref{eq:LLscaling}.  For example, electric and magnetic dipoles both exhibit $1/v_D^4$, but the electric dipole emission is only in TT-modes. This scaling is only introduced because these couplings were studied in isolation, as shown below.

From the UV perspective, the multipole moments are not independent and the emission rate rather probes the entire vertex factor $i\Gamma^{\nu\alpha\beta}$, so the inferences among different operators should enter. Based on their $C$ and $P$ properties, the operators can be grouped in four categories, as listed in Tab.~\ref{tab:scalingUV}. At the order of $O(m_N^0)$, only $\kappa_\Lambda$ and $\tilde \kappa_\Lambda$ are non-zero, corresponding to  $\mu_V = -Q_V m_V/2\neq 0$,  $d_V = -\tilde Q_V m_V/2\neq 0$, and $g_i^A =0$  ($i=1,4,5$) in the language of our effective interactions. Such relations among couplings, motivated by the UV theory, resolve the issue of the scaling in the last paragraph. For example, consider the first group in this table: the emission rates $\dot  Q_{\rm LL} $ and $\dot  Q_{\rm TL}$ are induced at the first order and indeed have the scaling introduced in Eq.~\eqref{eq:polscaling} at small-$m_V$ limit.  At the next-to-leading order $O(m_N^{-2})$, $\lambda_\Lambda$ and $g_1^A$ become non-zero too, which open up the TT emission channel, with $\dot  Q_{\rm TT}\propto g_D^4 $ following Eq.~\eqref{eq:polscaling}.  The same scaling can also be observed for other groups in the dominant contribution of the emission rates, as summarized in  Tab.~\ref{tab:scalingUV}. 

The analysis above shows that, given an underlying UV model, the coefficients of the effective interactions can be connected. Taken together, they reproduce the scalings of Eq.~\eqref{eq:polscaling}. We therefore conclude that caution must be exercised when translating the constraints on the effective operators into bounds on a UV model, as there can exist significant destructive interferences in the squared amplitude calculation.  On the other hand,  from the perspective of dimensional analysis the function $f(s)$ of the considered operators scale as $s^3$,  $s^4$ and $s^5$. Therefore our stellar lower bounds can be easily re-scaled to constrain interactions with similar dependence on $s$ for dark state masses well below the stellar temperature. Take the example of a milli-charged $V$, for which $\dot  Q_{\rm LL}$ dominates and $f(s)\propto s^3$  in the limit of $m_V\ll \sqrt{s}$. Our stellar/freeze-in/BBN  results  on $Q_V$ then approximately as well apply to $\epsilon e /m_V^2$, where $\epsilon$ is the milli-charge of $V$.

Finally, we also provide some comments on the naturalness of the smallness of $m_V$. 
Naive dimensional analysis together with the requirement of the masslessness of $V$ in the limit of unbroken SU(2)$_D$ suggests that  the radiative correction  to the mass of $V$ scales as $\delta m_V^2 \sim m_V^2 \,g_D^2/16\pi^2$ or $m_V^2 \,y^2/16\pi^2$, where $y$ denotes the Yukawa coupling of intermediate fermions. Requiring that this correction does not  exceed $m_V$, \emph{i.e.}, $\delta m_V/m_V \lesssim {\mathcal O}(1)$, suggests perturbative gauge and Yukawa couplings, and yields a consistency constraint on the size of the effective coupling. 
For the magnetic (electric) dipole one obtains  $\mu_V (d_V) m_V \sim g_D^2 /16\pi^2 \lesssim {\mathcal O}(1)$, 
which is respected in the entire parameter space explored in Fig.~\ref{fig:stellar_bound27}.  Similarly, we obtain $Q_V (\tilde Q_V)\lesssim 10^8\ \GeV^{-2} (\MeV/m_V)^2$,  which is a very mild requirement on the high mass end. 
Finally, the remaining couplings are constrained by $g_i^A/m_V^2\lesssim 10^{-4}(\TeV/m_N)^2$. 
It should be noted, however, that in the concretely considered UV model with heavy integer charged fermions $\Psi_{E}^i$ and $\Psi_N$,  constraints from the Large Hadron Collider (LHC) imply a fermion mass scale of several hundred of GeV. This predicts much smaller values of $g_i^A$ than are being probed in Figs.~\ref{fig:stellar_bound1} and~\ref{fig:stellar_bound27}. In summary, the considered UV completion helped us to answer important questions of the $m_V\to 0$ limit, but is not capable to populate the presented mass-coupling planes in their entirety;  we leave such model building challenges for future work.

\section{Conclusions}
\label{sec:conclusions}

In this work, we consider the neutral complex vector particles~$V^\mu$ below the GeV mass scale which are electrically neutral but share a coupling to the SM photon through higher-dimensional multipole moments. We study magnetic and electric dipole ($\mu_V$ and $d_V$), electric and magnetic quadrupole ($Q_V$ and $\tilde Q_V$) interactions, the anapole ($g_5^A$) and a CP-odd toroidal ($g_4^A$) moment, as well as a charge radius ($g_1^A$) interaction. Together, they make the complete list of seven electromagnetic moments a neutral vector particle can possess. 

We compute the relic density from freeze-out and freeze-in and contrast the predictions with the most important  astrophysical and cosmological constraints for each of the interaction, treating them as formally independent Wilson coefficients.
A thermal DM candidate $V$ from freeze-out requires $m_V \gtrsim 1\,\MeV$ so that annihilation into electron pairs becomes efficient. We then find that the combination of direct and indirect detection constraints as well as limits from cosmic ray physics exclude all but the toroidal or anapole moment as the origin for the SM origin. In contrast, freeze-in is possible for any mass considered (keV to GeV range). Because of the mass-dimensionality of  the effective operators, the relic abundance prediction depends on the reheating temperature. We find that for $T_{\rm RH} > 100\ \MeV$, the DM line is not touched by any of considered observables, and freeze-in  $V$-DM remains a valid but untested possibility. 

Independently from the DM hypothesis, the existence of (potentially unstable) $V$ particles with such couplings is probed by stellar energy loss arguments.  Specifically, we compute the emission rate of $V$-pairs in the Sun, HB and RG stars as well as from the proto-neutron star from SN1987A. All relevant production channels are accounted for: plasmon decay, Compton and Bremsstrahlung production, and, for the SN1987A constraint, electron-positron annihilation as well. A broad parameter region, principally below the MeV vector mass scale, is excluded from these considerations as well as from~BBN. 

A most important question regards the scaling of constraints with diminishing vector mass. 
We clarify the validity of the obtained results by explicit reference to a UV model that induces six of the seven operators. The calculations are valid as long as the invariant di-vector mass satisfies $\sqrt{s}\ll v_D$ and the emission rates remain finite for $m_V\to 0$. However, a UV model also connects various operators. We show that they are grouped according to their $C$ and $P$ transformation property, and that when the interactions, within such group, are jointly taken into account,  the stellar emission rates into the various polarization states exhibit the correct scaling as naive dimensional analysis suggests. 

How ``dark'' DM and physics beyond SM in general need to be is a question that finds a quantifiable and systematic answer by constraining the various coefficients of the  vertex function with the photon. In this work, we provide this answer of a dark vector particle with a mass below the GeV-scale. 

\vspace{0.2cm}

\paragraph*{Acknowledgments.}
We thank Ryo Nagai for collaboration in the initial stages of this project.
This work was supported by the Austrian Science Fund FWF: FG-1N (Research Group); the U.S.~National Science Foundation (NSF) Theoretical Physics Program, Grant PHY-1915005; the Research Network Quantum Aspects of Spacetime (TURIS); the Collaborative Research Center SFB1258; the Deutsche Forschungsgemeinschaft (DFG, German Research Foundation) under Germany's Excellence Strategy - EXC-2094 - 390783311; JSPS Grant-in-Aid for Scientific Research Grant No.20H01895; World Premier International Research Center Initiative (WPI Initiative), MEXT, Japan; JSPS Core-to-Core Program Grant No. JPJSCCA20200002. Funded/Co-funded by the European Union (ERC, NLO-DM, 101044443). Views and opinions expressed are however those of the author(s) only and do not necessarily reflect those of the European Union or the European Research Council. Neither the European Union nor the granting authority can be held responsible for them.

~\\

\appendix

\section{Couplings from the UV}
\label{app:uvcouplings}
Here we provide the explicit expressions for the various moment-interactions of the explicit UV model considered in~\cite{Hisano:2020qkq} and discussed in Sec.~\ref{sec:UV} where we take the limit that the vector $V$ is much lighter than the charged fermions generating the effective interactions,%
\footnote{
The limiting expressions for $\mu_V$, $Q_V$, and $g_1^A$ correct a typo in~\cite{Hisano:2020qkq}, where the factors $(1/r_N^2 - 1/r_{E_i}^2)$ in Eqs.~(4.13-4.15) there should have read $(r_N^2 - r_{E_i}^2)$. Here $r_{N, E_i} = m_{N, E_i}/m_V$. 
We thank R.~Nagai for this point. 
}
\begin{widetext}
\begin{align}
\mu_V & =-e \frac{g_D^2}{64 \pi^2} \frac{1}{m_{V}} \sum_{i=1}^2 (1-x_i^2)\left[\left(\left|\left(V_L\right)_{1 i}^2\right|^2+\left|\left(V_R\right)_{1 i}^2\right|^2\right) \mathcal{G}_\mu^{(1)}\left(x_i\right)+2 \operatorname{Re}\left(\left(V_L\right)_{1 i}^{\star}\left(V_R\right)_{1 i}^{\star}\right) \mathcal{G}_\mu^{(2)}\left(x_i\right)\right] \,,\\
d_V & =e \frac{g_D^2}{64 \pi^2} \frac{1}{m_V}  \sum_{i=1}^2 \operatorname{Im}\left(\left(V_L\right)_{1 i}^{\star}\left(V_R\right)_{1 i}^{\star}\right)\,\mathcal{G}_d^{(1)}\left(x_i\right) \,, \\
Q_V & =-e \frac{g_D^2}{64 \pi^2} \frac{1}{m_{V}^2} \sum_{i=1}^2 (1-x_i^2) \left[\left(\left|\left(V_L\right)_{1 i}^2\right|^2+\left|\left(V_R\right)_{1 i}^2\right|^2\right) \mathcal{G}_Q^{(1)}\left(x_i\right)+2 \operatorname{Re}\left(\left(V_L\right)_{1 i}^{\star}\left(V_R\right)_{1 i}^{\star}\right) \mathcal{G}_Q^{(2)}\left(x_i\right)\right] \,,\\
\tilde{Q}_V & =e \frac{g_D^2}{64 \pi^2}  \frac{1}{m_{V}^2}  \sum_{i=1}^2 \operatorname{Im}\left(\left(V_L\right)_{1 i}^{\star}\left(V_R\right)_{1 i}^{\star}\right)(-2) \mathcal{G}_d^{(1)}\left(x_i\right)  \,,\\
g_1^A & =-\frac{g_D^2}{64 \pi^2} \frac{m_{V}^2}{m_N^2} \sum_{i=1}^2 (1-x_i^2) \left[\left(\left|\left(V_L\right)_{1 i}^2\right|^2+\left|\left(V_R\right)_{1 i}^2\right|^2\right) \mathcal{G}_1^{(1)}\left(x_i\right)+2 \operatorname{Re}\left(\left(V_L\right)_{1 i}^{\star}\left(V_R\right)_{1 i}^{\star}\right) \mathcal{G}_1^{(2)}\left(x_i\right)\right]
\,,\\
g_5^A & =\frac{g_D^2}{128 \pi^2} \frac{m_{V}^2}{m_N^2}  \sum_{i=1}^2\left(\left|\left(V_L\right)_{1 i}^2\right|^2-\left|\left(V_R\right)_{1 i}^2\right|^2\right) \mathcal{G}_5\left(x_i\right) \,,
\end{align}
\end{widetext}
If the kinetic mixing between the photon and the third component of dark SU(2) gauge group, $W_D^0$,  is generated, a correction $2 ({g_D}/{e}) \epsilon  /(s-m^2_{W_D^0})$ is added to $g_1^A/m_V^2$. Thus its contribution to $g_1^A$ is approximately proportional to  $m_V^2$ at the limit of $s \gg m^2_{W_D^0}$.  Here, $x_i = m_{E_i}/m_N$ and the loop functions are given by 
\begin{align}
\mathcal{G}_\mu^{(1)}(x) & =\frac{\left(1-x^4+4 x^2 \log (x)\right)}{\left(1-x^2\right)^2} \,,\\
\mathcal{G}_\mu^{(2)}(x) & =-\frac{4 x\left(1-x^2+\left(1+x^2\right) \log (x)\right)}{\left(1-x^2\right)^2}\,, \\
\mathcal{G}_Q^{(1)}(x) & =-\frac{2\left(1-x^4+4 x^2 \log (x)\right)}{\left(1-x^2\right)^2} \,,\\
\mathcal{G}_Q^{(2)}(x) & =\frac{8 x\left(1-x^2+\left(1+x^2\right) \log (x)\right)}{\left(1-x^2\right)^2} \,,\\
\begin{split}
\mathcal{G}_1^{(1)}(x) & =-\frac{1}{9\left(1-x^2\right)^4}\left[2(11-45 x^2+45 x^4-11 x^6  \right.  \\  &\quad \left.+12\left(1-2 x^2-2 x^4+x^6\right) \log (x))\right] 
\end{split}\,,\\
\mathcal{G}_1^{(2)}(x) & =-\frac{x\left(12\left(1-x^4\right)+8\left(1+4 x^2+x^4\right) \log (x)\right)}{3\left(1-x^2\right)^4} \,,\\
\mathcal{G}_d^{(1)}(x) & =-\frac{8 x \log (x)}{\left(1-x^2\right)} \,,\\
\mathcal{G}_d^{(2)}(x) & =-\frac{8 x\left(1-x^2+\left(1+x^2\right) \log (x)\right)}{\left(1-x^2\right)^3} \,,\\
\mathcal{G}_5(x) & =\frac{2\left(3\left(1-x^4\right)+4\left(1+x^2+x^4\right) \log (x)\right)}{3\left(1-x^2\right)^3}\,. 
\end{align}
These loop functions are valid up to the order of $\mathcal{O}(1/m_N^2)$. The kinetic mixing between photon and $W_D^0$ is given by
\begin{align}
\epsilon&= -\frac{eg_D}{12\pi^2}\left[2 \log{x_2}+ \left(|(V_R)_{11}|^2+|(V_L)_{11}|^2\right)\log\frac{x_1}{x_2}\right]\,.
\end{align}
One observes that $2\mathcal{G}_\mu^{(1)}(x) + \mathcal{G}_Q^{(1)}(x) =0$ and  $2\mathcal{G}_\mu^{(2)}(x) + \mathcal{G}_Q^{(2)}(x) =0$, suggesting at the  first order, $2\mu_V + m_V Q_V =0$. That is, $\lambda_\Lambda$ only appears at the  order of $\mathcal{O}(1/m_N^2)$, for which we have calculated to yield 
\begin{widetext}
\begin{align}
\lambda_\LambdaUV    & = {\LambdaUV^2 \over e m_V}\left( \mu_V + {m_V Q_V \over 2}   \right) \notag\\
&= - \frac{g_D^2}{32 \pi^2 } \frac{\LambdaUV^2}{m_N^2} \sum_{i=1}^2 (1-x_i^2)\left(\left|\left(V_L\right)_{1 i}^2\right|^2+\left|\left(V_R\right)_{1 i}^2\right|^2\right)\left[ {   -1 + x_i^2 \left(-9 + 9 x_i^2 + x_i^4 -   12 (1 + x_i^2) \log(x_i)\right)\over 9 (-1 + x_i^2)^5  } \right]  \,.
\end{align}
\end{widetext}
In contrast, in this UV model there is always $2d _V + m_V \tilde Q_V =0$, and thus $\tilde \lambda_\Lambda  =0$,  at one-loop level.

If we fix the dimensionless coefficient $x_i$ and the dark symmetry breaking scale, denoted as ${v}_D \equiv 2m_V/g_D$, for non-degenerate fermion mass we obtain the effective coefficients from dimensional analysis (and the consequent stellar luminosity scaling for $T\gg m_V$) in Tab.~\ref{tab:scaling} of the main text.

\bibliography{refs}

\end{document}